\documentclass[10pt, a4paper]{article}
\pdfoutput=1 
\usepackage{jcappub}
\usepackage{graphicx}
\usepackage{amsmath,amssymb}
\usepackage{bm}
\usepackage{color}
\usepackage{tensor}
\usepackage{hyperref}
\usepackage{slashed}
\usepackage{multirow}

\def\be{\begin{equation}}
\def\ee{\end{equation}}
\def\ba{\begin{eqnarray}}
\def\ea{\end{eqnarray}}
\def\nn{\nonumber}

\def\GeV{\rm \, GeV}

\def\eV{\rm \, eV}
\def\cm{\rm \, cm}

\newcommand{\meV}{\, {\rm meV}}
\newcommand{\keV}{\, {\rm keV}}
\newcommand{\MeV}{\, {\rm MeV}}

\newcommand{\yr}{\, {\rm yr}}

\newcommand{\per}{\, .}
\newcommand{\com}{\, ,}
\newcommand{\eref}[1]{Eq.~(\ref{#1})}
\newcommand{\aref}[1]{Appendix~\ref{#1}}
\newcommand{\eg}{{\it e.g.}}
\newcommand{\ie}{{\it i.e.}}

\newcommand{\abs}[1]{\left| #1 \right|}

\newcommand{\erefs}[2]{Eqs.~(\ref{#1})~and~(\ref{#2})}
\newcommand{\fref}[1]{Fig.~\ref{#1}}
\newcommand{\sref}[1]{Sec.~\ref{#1}}
\newcommand{\tref}[1]{Table~\ref{#1}}
\newcommand{\rref}[1]{Ref.~\cite{#1}}
\newcommand{\tritium}{\tensor[^3]{\rm H}{}}
\newcommand{\Hethree}{\tensor[^3]{\rm He}{}}
\newcommand{\rSN}{r_{\text{\sc sn}}}
\newcommand{\CNB}{\text{\sc c}\nu\text{\sc b}}
\def\nue{\mathrel{{\nu_e}}}
\def\numu{\mathrel{{\nu_\mu}}}
\def\nutau{\mathrel{{\nu_\tau}}}

\def\barnue{\mathrel{{\bar \nu}_e}}
\def\barnumu{\mathrel{{\bar \nu}_\mu}}
\def\barnutau{\mathrel{{\bar \nu}_\tau}}

\def\t13{\mathrel{{\theta_{13}}}}
\def\y12{\mathrel{{\tan^2 \theta_{12}}}}
\def\c2{\mathrel{{\chi^2 }}}

\newcommand{\n}{neutrino}
\newcommand{\ns}{neutrinos}

\newcommand{\cnb}{C$\nu$B}

\def \gta {\mathrel{\vcenter{\hbox{$>$}\nointerlineskip\hbox{$\sim$}}}}

\def\nue{\mathrel{{\nu_e}}}

\def\numu{\mathrel{{\nu_\mu}}}
\def\nutau{\mathrel{{\nu_\tau}}}

\def\barnue{\mathrel{{\bar \nu}_e}}
\def\barnumu{\mathrel{{\bar \nu}_\mu}}
\def\barnutau{\mathrel{{\bar \nu}_\tau}}

\title{Detecting non-relativistic cosmic neutrinos by capture on tritium: phenomenology and physics potential}

\date{\today}

\author[]{Andrew J. Long,}
\author[]{Cecilia Lunardini,}
\author[]{Eray Sabancilar} 

\affiliation[]{ Physics Department, Arizona State University, Tempe, Arizona 85287, USA.}

\emailAdd{andrewjlong@asu.edu}
\emailAdd{Cecilia.Lunardini@asu.edu}
\emailAdd{Eray.Sabancilar@asu.edu}

\abstract{
We study the physics potential of the detection of the Cosmic Neutrino Background via neutrino capture on tritium, taking the proposed PTOLEMY experiment as a case study. With the projected energy resolution of $\Delta \sim$ 0.15 eV, the experiment will be sensitive to neutrino masses with degenerate spectrum, $m_1 \simeq m_2 \simeq m_3 = m_\nu \gtrsim 0.1$ eV. These neutrinos are non-relativistic today; detecting them would be a unique opportunity to probe this unexplored kinematical regime. The signature of neutrino capture is a peak in the electron spectrum that is displaced by $2 m_{\nu}$ above the beta decay endpoint. The signal would exceed the background from beta decay if the energy resolution is $\Delta \lesssim  0.7~ m_\nu $.  Interestingly, the total capture rate depends on the origin of the neutrino mass, being $\Gamma^{\rm D} \simeq 4$ and $\Gamma^{\rm M} \simeq 8$ events per year (for a 100 g tritium target) for unclustered Dirac and  Majorana neutrinos, respectively. An enhancement of the rate of up to $\mathcal{O}(1)$ is expected due to gravitational clustering, with the unique potential to probe the local overdensity of neutrinos. Turning to more exotic neutrino physics, PTOLEMY could be sensitive to a lepton asymmetry, and reveal the eV-scale sterile neutrino that is favored by short baseline oscillation searches. The experiment would also be sensitive to a neutrino lifetime on the order of the age of the universe and break the degeneracy between neutrino mass and lifetime which affects existing bounds. 
}

\begin{document}
\maketitle
\flushbottom


\section{Introduction}
\label{sec:intro}

The Cosmic Neutrino Background (\cnb) is a cardinal
feature of early universe cosmology, and holds the key to understanding many of its most interesting and well-studied phenomena: from the primordial synthesis of elements, to the anisotropies of the cosmic microwave background (CMB), and even to the formation of dark matter halos (for a review see, \eg, \cite{Dolgov:2002wy,Lesgourgues:2006nd,Quigg:2008ab,Dolgov:2008hz}). 

The body of information from cosmological probes, on the composition and distribution of matter and energy in the early universe, constitutes a very strong indirect evidence that the \cnb\ exists and confirms the Standard Model's prediction of its energy density.  Specifically, measurements of the CMB anisotropies and the large scale distribution of galaxies have already supplied two key pieces of data:  a measurement of the effective number of neutrino species, $N_{\rm eff}$, and a strikingly strong upper bound on the sum of the \n\ masses, $\sum m_{\nu}$.  
The most recent values from the Planck satellite read as follows \cite{Ade:2013zuv}:  
\be
N_{\rm eff} = 3.30 \pm 0.27 
\qquad {\rm and} \qquad
\sum m_{\nu} < 0.23 \eV~~\text{at}~95\%~\text{CL} \per
\label{cosmobound}
\ee
With the next generation of CMB telescopes, the sensitivity to $\sum m_{\nu}$ will be reduced to the $0.05 \eV$ level, which could allow for a measurement \cite{Bock:2009xw}.   

At this time, however, we still lack the truly golden signature of the \cnb\ that only a laboratory-controlled, direct detection experiment could provide.  Such a detection would not only complement other cosmological probes, and thereby help to resolve degeneracies among the neutrino model parameters, but it would access a whole array of phenomena that are beyond the reach of cosmological measurements. In the first place, a direct detection would confirm that the relic \ns\ are still present in the universe today -- a reasonable assumption if the neutrinos are stable, but one which has no empirical confirmation from cosmological observations alone. To put this less dramatically, a direct detection of the \cnb\ would probe late time effects, those occurring after recombination, such as \n\ clustering (and therefore the \n\ coupling to gravity), changes in the \cnb\ flavor composition or number density due to \n\ decay, or decay of heavy relics into \ns, and so on.  Perhaps even more importantly, a direct detection of the \cnb\ would constitute the first probe of {\it non-relativistic} neutrinos (since current detectors are only sensitive to relatively large neutrino masses), and thereby open the window onto an entirely new kinematical regime.  Studying non-relativistic neutrinos could allow for tests of certain neutrino properties that are difficult to access at high momentum such as the Dirac or Majorana character of neutrinos.
\begin{figure}[t]
\begin{center}
\includegraphics[width=0.65\textwidth]{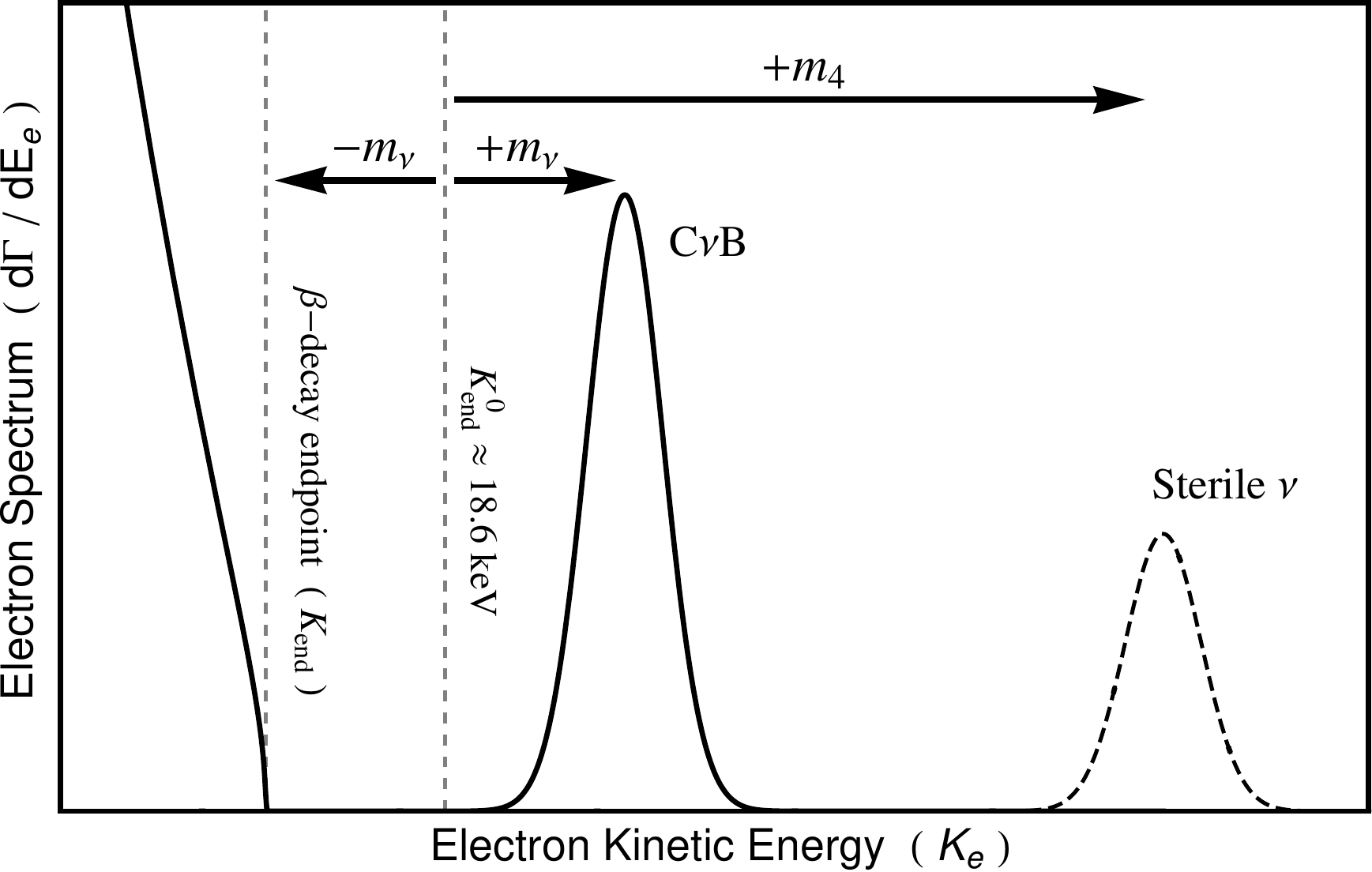} 
\caption{\label{fig:cartoon}
A cartoon illustrating the expected signal from the three active \cnb\ neutrinos of mass $m_1 \simeq m_2 \simeq m_3 = m_\nu$  (solid line), and from a hypothetical, mostly sterile, \n\ mass state, $\nu_4$, of mass $m_4$  (dashed line).  
The \cnb\ signal is displaced from the beta decay endpoint by $2 m_{\nu}$,  and the $\nu_4$ signal would be displaced by $m_{\nu} + m_{4}$.  
The signal and background are not represented to scale.  
Here $K_e = E_e - m_e$ is the electron kinetic energy, and $K^{0}_{\rm end}$ denoted by the vertical dashed line refers to the beta decay end point kinetic energy in the $m_\nu = 0$ limit. For a details, see Sec.~\ref{sec:cnb_detection} and Sec.~\ref{sec:ptolemy}.   
}
\end{center}
\end{figure}

Given the importance of a direct detection of the \cnb, it is not surprising that research in this field has been active and uninterrupted. In 1962 Weinberg was the first to advocate for \cnb\ detection via neutrino capture on beta-decaying nuclei (NCB) since this process requires no threshold energy \cite{Weinberg:1962zza}. The NCB technique is primarily limited by availability of the target material and by the need for extremely high precision in measuring the electron energy\footnote{In his paper, Weinberg reports of an experimental attempt being carried out by R.~W.~P.~Drever at the University of Glasgow at the time of his writing, resulting in a preliminary bound on the \cnb\ Fermi energy $E_F < 500 \eV$. We have been unable to retrieve any other information on this early experiment.  }.  
Other detection methods have their own challenges.  The Stodolsky effect, for instance, could allow \cnb\ neutrinos to be detected by their coherent scattering on a torsion balance \cite{Stodolsky:1974aq,Shvartsman:1982sn}, but the expected accelerations are well below the sensitivity of current detectors \cite{Duda:2001hd,Hedman:2013hha}, and vanishes if the \cnb\ is lepton-symmetric. In the last few years, attention has focused again on Weinberg's NCB technique, and a number of detailed studies have assessed the prospects for detection with a tritium target \cite{Cocco:2007za,Lazauskas:2007da,Blennow:2008fh,Li:2010sn,Faessler:2011qj}. In this type of an experiment, the smoking gun signature of \cnb\ capture, $\nu + \tensor[^3]{\rm H}{} \to \tensor[^3]{\rm He}{} + e^{-}$, is a peak in the electron spectrum at an energy of $2m_{\nu}$ above the beta decay endpoint; see \fref{fig:cartoon}. Detecting this peak requires an energy resolution below the level of $m_{\nu} = O(0.1 \eV)$.  Compared to other beta-decaying nuclei, tritium makes a particularly attractive candidate target because of its availability, high neutrino capture cross section, long lifetime (12 years), and low Q-value \cite{Cocco:2007za}. For a 100 gram target, the expected capture rate is approximately 10 events per year \cite{Cocco:2007za}. So far, however, difficulties in achieving the necessary sub-eV energy resolution, and in controlling broadening of the electron energy distribution have precluded any serious experimental effort.  

In 2012/2013 the Princeton Tritium Observatory for Light, Early-Universe, Massive-Neutrino Yield (PTOLEMY), located at the Princeton Plasma Physics Laboratory, began developing a technology that could help to solve the energy resolution challenges \cite{Betts:2013uya}.  The tritium nuclei will be deposited onto a source disk, such as a graphene substrate.  This geometry helps to reduce electron backscatter, and thereby achieve an energy resolution of $\Delta \sim 0.15 \eV$, of the order of the neutrino mass scale.  With this resolution and a 100 gram sample of tritium, PTOLEMY could transform \cnb\ detection from fantasy into reality.  

These recent advances, and especially the prospect of an having an experimental search in the near future, motivate studying the phenomenology of NCB in more detail. This is the spirit of our paper. In particular, the main novelties of our study are the sensitivity to the Dirac or Majorana nature of the \n, a more detailed analysis of the background rate, and the potential of the NCB to study a number of effects ranging from expected standard phenomenology, such as gravitational clustering and mass hierarchy, to more exotic ideas like lepton asymmetry, sterile \ns, \n\ decay and non-standard thermal history. 

The plan of the paper is as follows. In Sec.~\ref{sec:nucap}, we discuss the creation and evolution of the \cnb\ \ns, and calculate the polarized neutrino capture cross section and the capture rate for tritium nucleus to clarify the difference between the Dirac and Majorana neutrinos. A detailed calculation of the neutrino capture kinematics and the polarized neutrino scattering amplitude is given in Appendix~\ref{app:nucapture}. In Sec.~\ref{sec:ptolemy}, we focus on a PTOLEMY-like experiment, and treat the tritium beta decay as the main background for the tritium neutrino capture signal. In particular, we study the signal to noise ratio by taking into account the finite energy resolution of the detector, and find the required energy resolution for various neutrino masses. In Sec.~\ref{sec:outcome}, we discuss the difference between the Dirac and the Majorana neutrinos, the effect of the mass hierarchy, and gravitational clustering of neutrinos. In Sec.~\ref{sec:sterile}, we discuss the sensitivity to an eV (and sub-eV) scale sterile neutrino and a keV-scale warm dark matter sterile neutrino. In Sec.~\ref{sec:exotica}, we discuss various effects of new physics that can lead to an enhancement or suppression of the \cnb\ number density, such as lepton asymmetry in the neutrino sector, neutrino decay, and late time entropy injection. A summary and discussion follow in Sec.~\ref{sec:disc}.


\section{Cosmic background neutrinos and their capture on tritum}
\label{sec:nucap}

In this section we will trace the history of a \cnb\ neutrino, considering its production, propagation and detection. In reviewing the physics of these, we  emphasize two critical points:  the distinction between Dirac and Majorana neutrinos and the distinction between helicity and chirality. These are important to derive one of the main conclusions, namely that the \cnb\ capture rate for Dirac and Majorana neutrinos differ by a factor of 2.

\subsection{Thermal history of the \cnb}
\label{sec:cnb_creation}

Let us first discuss the production of \ns\ in the early universe, \ie, their properties up to the point when they start free streaming. In the hot, dense conditions of the early universe, the neutrinos maintained thermal equilibrium with the plasma (electrons, positrons, and photons) through scattering processes such as 
\begin{align}\label{eq:thermal_processes}
	\nu e \longleftrightarrow \nu e
	\qquad {\rm and} \qquad
	e^+ e^- \longleftrightarrow \nu \bar{\nu} \per
\end{align}
These processes are mediated by the weak interaction, therefore the \ns\ are produced as flavor eigenstates, $\nue,\numu,\nutau, \barnue, \barnumu, \barnutau$. The scattering rate of the processes in \eref{eq:thermal_processes} depends strongly on the temperature $T$, as $\Gamma \approx G_F^2 T^5$, where $G_F \approx 1.2 \times 10^{-5} \GeV^{-2}$ is the Fermi constant. At this time the spectrum of the \ns\ is thermal, given by the Fermi-Dirac distribution, $f_{\rm FD}({\bf p}, T) =(1 + e^{E / T})^{-1}$, where $E = \sqrt{{\bf p}^2 + m_{\nu}^2}$ and $T$ is the temperature of the plasma. Integrating over the phase space gives the number density of \ns\ per degree of freedom (flavor and spin):
\begin{align}\label{eq:n_fo}
	n_{\nu} (T) = \frac{3 \zeta(3)}{4\pi^2} T^3 \per
\end{align}
(We will neglect the possibility of a lepton asymmetry for now, and return to this point in \sref{sub:lepton_asym}.)  

At a temperature of $T_{\rm fo} \sim \MeV$, the scattering rate dropped below the Hubble expansion rate, $H \approx T^2 / M_P$ (where $M_P \approx 2.4 \times 10^{18} \GeV$), and as a consequence the neutrinos fell out of thermal equilibrium  (``freeze out").  
Effectively, the time of freeze out can be considered as the instant of production of the \cnb\ \ns\ that we hope to detect today, since after this time the \ns\ simply free stream.  
In any case, it is easy to recognize that our conclusions do not depend on the exact instant of production of each \n. 

Between freeze out and the present epoch, \ns\ undergo a number of interesting effects, that we summarize below. \\

\noindent
{\it (i) redshift.}\\
In the sudden freeze out approximation, the phase space distribution function after decoupling is given by an appropriate redshifting of the distribution function that was realized at decoupling. This leads to a modified Fermi-Dirac distribution\footnote{This approximation agrees with exact solutions of the Boltzmann equation to within $O(0.2\%)$ \cite{Gnedin:1997vn}.  }\footnote{Note that  Eq. (\ref{eq:f_fo})  is valid for any value of $p$ and of the \n\ mass.  In it, the 
mass term is suppressed by a factor of $( 1 + z ) / (1 + z_{\rm fo} ) \ll 1$, which we neglect.}
\begin{align}\label{eq:f_fo}
	f_{\nu}[p(z) \, , \,T_\nu(z)] = \frac{1}{ e^{p(z) / T_{\nu}(z)} + 1 } \com \qquad dn_\nu = \frac{d^3 p(z)}{(2\pi)^3} f_{\nu}[p(z) \, , \,T_\nu(z)]  \com
\end{align}
where
\begin{align}\label{eq:Tnu_of_t}
		p(z) =\frac{1 + z}{1 + z_{\rm fo}} p_{\rm fo} \com \qquad T_{\nu}(z) =\frac{1 + z}{1 + z_{\rm fo}} T_{\rm fo} 
\end{align}
are the neutrino momentum and the effective neutrino temperature, respectively.   Here they are expressed in terms of the momentum variable $p_{\rm fo}$, the neutrino temperature and redshift at freeze out, $T_{\rm fo}$ and $z_{\rm fo} \simeq 6 \times 10^{10}$.

After neutrino freeze out, the \cnb\ relic abundance is given by \eref{eq:n_fo}, where \eref{eq:Tnu_of_t} gives the effective neutrino temperature.  
As the universe expands, $z$ decreases and so too does $T_{\nu}$.  
Meanwhile the photons redshift like 
\begin{align}\label{eq:Tg_of_t}
	T_{\gamma}(z) = \frac{1 + z}{1 + z_{\rm fo}} \frac{g_{\ast}(z_{\rm fo})^{1/3} }{ g_{\ast}(z)^{1/3} } T_{\rm fo} \com
\end{align}
where $g_{\ast}(z) = 45 s(z) / [2 \pi^2 T(z)^3]$ and $s(z)$ is the entropy density at epoch $z$.  
After electron-positron annihilation freezes out at $T \approx 100 \keV$, this entropy is transferred to the photons, which causes them to cool less quickly.  This leaves the \cnb\ at a relatively lower temperature, 
\begin{align}\label{eq:Tnu}
	T_{\nu} \approx (4/11)^{1/3} T_{\gamma} \per
\end{align}
We can extrapolate until today when the temperature of the CMB is measured to be $T_{\gamma} = 0.235 \meV$ \cite{Ade:2013zuv}. Then, the relationship above predicts the current temperature of the \cnb\ to be $T_{\nu} = 0.168 \meV$.  Using \eref{eq:n_fo} this corresponds to a number density of 
\begin{align}\label{nz}
	n_\nu (z) =  n_0 (1+z)^3,
\end{align}
where
\begin{align}\label{eq:n0_def}
	n_{0} \approx 56 \, {\rm cm}^{-3}
\end{align}
per degree of freedom or $6 n_0 \approx 336 \, {\rm cm}^{-3}$ for the entire \cnb.  
Using \eref{eq:f_fo}, the root mean square momentum of neutrinos in the present epoch can be found to be
\begin{align}\label{eq:pbar0}
	\overline{p}_0 \approx 0.603 \meV \per
\end{align}
Since we are only interested in $m_\nu \gtrsim 0.1 \eV$ for the direct detection purposes, and $\overline{p}_{0} \ll m_{\nu} \sim 0.1 \eV$, we assume that the \cnb\ neutrinos are extremely non-relativistic today.\\

\noindent
{\it (ii) quantum decoherence.}\\
As previously mentioned, \ns\ are produced as flavor eigenstates, $\nu_\alpha$, which are a coherent superposition of mass eigenstates, $\nu_i$: $\nu_\alpha = \sum_i U_{\alpha i}~ \nu_i$, with $U$ being the Pontecorvo-Maki-Nakagawa-Sakata (PMNS) matrix \cite{Pontecorvo:1957cp,Pontecorvo:1967fh,Maki:1962mu} probed by oscillation experiments. 

Over time, the neutrino wavepacket decoheres as the different mass eigenstates $\nu_i$ propagate at different velocities \cite{Akhmedov:2009rb}.  
The timescale for this decoherence, $\Delta t$, can be estimated by solving $(v_1 - v_2) \Delta t \approx \lambda$ where $v_{i} \approx p / \sqrt{p^2 + m_i^2} \approx 1 - m_i^2 / 2 p^2$
are the velocities of two mass eigenstates and $\lambda \approx p^{-1}$ is the Compton wavelength of the wavepacket.  The solution for $\Delta t$, in units of Hubble time ($H^{-1} \approx M_P / T^2$), is:
\begin{align}
	\frac{\Delta t}{H^{-1}} \approx \frac{2 p}{m_2^2 - m_1^2} \frac{T^2}{M_P}  \approx 10^{-7}~  \com
\end{align}
where we used $m_2 \approx 2 m_1 \approx 0.1 \eV$ and $p \approx T_{\rm fo} \approx 1 \MeV$.
It is found that the flavor eigenstate \cnb\ neutrinos quickly decohere into their mass eigenstates on a time scale much less than one Hubble time \cite{Eberle:2004ua}.  Since we do not expect the decoherence to affect the relative abundances, we then conclude that \ns\ with the mass values of interest here, are present in the universe today as mass eigenstates, equally populated with an abundance given by \eref{eq:n_fo}.  \\

\subsection{Helicity composition of the \cnb}
\label{sec:cnb_evolution}

Next, let us turn to the question of the neutrino spin state at production.   Recall that a field's chirality determines its transformation property under the Lorentz group, and that the weak interaction is chiral in nature, \eg, the left-chiral component of the electron interacts with the weak bosons, but the right-chiral component does not.  Therefore \ns\ (anti-neutrinos) are only produced in the left-chiral (right-chiral) state. Chirality should not be confused with a particle's helicity, which is given by the projection of its momentum vector onto its spin vector.  

Since the \cnb\ neutrinos are ultra-relativistic at freeze out ($T_{\rm fo} \gg m_{\nu}$), we do not (yet) need to explicitly distinguish helicity and chirality, which exactly coincide for massless particles.  For simplicity, here we will use the terminology ``left-handed'' to refer to a relativistic state that is left-helical and left-chiral, and we do similarly with the right-handed states.  

At this point is it convenient to enumerate all possible spin states.  If the neutrinos are Dirac particles then we have four degrees of freedom per generation, which we will label as 
\begin{align}\label{eq:D_state}
\begin{array}{lcl}
	\nu_{L} & \quad & \text{left-handed active neutrino} \\
	\bar{\nu}_{R} & \quad & \text{right-handed active anti-neutrino} \\
	\nu_{R} & \quad & \text{right-handed sterile neutrino} \\
	\bar{\nu}_{L} & \quad & \text{left-handed sterile anti-neutrino} 
\end{array} \per
\end{align}
Neutrinos and anti-neutrinos are distinguished by their lepton number, which is a conserved quantity.  The states  $\nu_L$ and $\bar{\nu}_R$ are active in the sense that they interact via the weak interaction, while in contrast  $\nu_R$ and $\bar{\nu}_L$ are labeled as sterile because they interact only via the Higgs boson (\ie, the mass term). This interaction is suppressed by a very small Yukawa coupling $y_{\nu} \approx m_{\nu} / v \approx 10^{-12}$, where $v \approx 246 \GeV$ is the vacuum expectation value of the Higgs field.  

The production mechanisms we have discussed above clearly apply only to the active states, which therefore acquire the abundance, $n_{\nu}(z)$, given by \eref{nz}.  Meanwhile, the sterile neutrinos can not come into thermal equilibrium with the SM, so it is reasonable to assume that their relic abundance is negligible compared to that of the active states\footnote{
One cannot exclude the possibility that there was a primordial abundance of sterile neutrinos, and to answer this question unambiguously one would have to specify the physics of the reheating phase that followed inflation. Nevertheless, it seems unlikely that this abundance was as large as $n_{\nu}(z)$ at the time of neutrino freeze out. As each of the SM fermion species froze out during the thermal history, they transferred their entropy to the remaining thermal species.  Each of these entropy injections would have diluted the decoupled sterile neutrinos.  (The physics is identical to the suppression of the \cnb\ abundance relative to the CMB abundance after $e^+ e^-$ annihilation.)
}. 
Then, for the Dirac case, we expect the spin state abundances to be
\begin{align}\label{eq:n_Dirac_at_fo}
\begin{array}{l}
	n(\nu_{L}) = n_{\nu}(z) \\
	n(\bar{\nu}_{R}) = n_{\nu}(z) \\
	n(\nu_{R}) \approx 0 \\
	n(\bar{\nu}_{L}) \approx 0 
\end{array} 
\end{align}
where $n_{\nu}(z)$ is given by \eref{nz}.  
The total \cnb\ abundance is given by $6 n_{\nu}(z)$ after summing over spin and flavor states.  

If the neutrinos are Majorana particles then lepton number is not a good quantum number, and we should avoid using the language ``neutrino'' and ``anti-neutrino''\footnote{Our language here differs from conventions in the literature.  When discussing Majorana neutrinos, it is customary to equate lepton number with chirality, such that the left-chiral particle is called a neutrino and the right-chiral particle is called an anti-neutrino. This language is very useful for discussing relativistic neutrinos, but impractical for  non-relativistic neutrinos, for which we must distinguish helicity and chirality.}. Instead, we will label the degrees of freedom as 
\begin{align}\label{eq:M_state}  
\begin{array}{lcl}
	\nu_{L} & \quad & \text{left-handed active neutrino} \\
	\nu_{R} & \quad & \text{right-handed active neutrino} \\
	N_{R} & \quad & \text{right-handed sterile neutrino} \\
	N_{L} & \quad & \text{left-handed sterile neutrino} 
\end{array} \per
\end{align}
As in the Dirac case, the active neutrinos interact weakly, and both the left- and right-handed states are populated at freeze out.  The sterile neutrinos interact only through the Higgs boson, like in the Dirac case, but now they are typically much heavier than even the electroweak scale (see, \eg, \cite{GellMann:1980vs, Mohapatra:1979ia, Yanagida:1980xy}).  As such, they will decay into a Higgs boson and a lepton, and their relic abundance today is zero.  
To summarize the Majorana case, we have 
\begin{align}\label{eq:n_Maj_at_fo}
\begin{array}{l}
	n(\nu_{L}) = n_{\nu}(z) \\
	n(\nu_{R}) = n_{\nu}(z) \\
	n(N_{R}) = 0 \\
	n(N_{L}) = 0 
\end{array} 
\end{align}
where once again the total \cnb\ abundance is $6 n_{\nu}(t)$.  

Let us discuss how the \n\ quantum states evolve starting from the composition at freezeout, Eqs.~(\ref{eq:n_Dirac_at_fo}) and (\ref{eq:n_Maj_at_fo}). To describe the cooling of \ns\ down to the present time, we need to abandon the ultrarelativistic approximation, and therefore study the regime where helicity and chirality do not coincide.  To do so, a key point to consider is that the helicity operator commutes with the free particle Hamiltonian, and its conservation is tied to the conservation of angular momentum. Instead, the chirality operator does not commute because of the mass term. Consequently, while the neutrinos are freely streaming, it is their helicity and not their chirality that is conserved \cite{Duda:2001hd}. Thus, we can determine the abundances today from Eqs.~(\ref{eq:n_Dirac_at_fo})~and~(\ref{eq:n_Maj_at_fo}) upon recognizing that ``handedness'' at freeze out translates into ``helicity'' today.  Let us denote $n(\nu_{h_L})$ as the number density of left-helical neutrinos, $n(\nu_{h_R})$ as the number density of right-helical neutrinos, and so on.  
Then the abundances today are, for Dirac \ns:
\begin{align}\label{eq:n_Dirac_today}
\begin{array}{l}
	n(\nu_{h_L}) = n_{0} \\
	n(\bar{\nu}_{h_R}) = n_{0} \\
	n(\nu_{h_R}) \approx 0 \\
	n(\bar{\nu}_{h_L}) \approx 0 
\end{array} 
\end{align}
and, for Majorana \ns:
\begin{align}\label{eq:n_Maj_today}
\begin{array}{l}
	n(\nu_{h_L}) = n_{0} \\
	n(\nu_{h_R}) = n_{0} \\
	n(N_{h_R}) = 0 \\
	n(N_{h_L}) = 0 
\end{array} 
\end{align}
where $n_0$ is given by \eref{eq:n0_def}.  Note that the total abundance is the same, $6 n_0$, in both cases. However, the \cnb\ contains both left- and right-helical active neutrinos in the Majorana case, but only left-helical active neutrinos in the Dirac case.    

Finally, we note that, if the neutrinos are not exactly free streaming, but instead they are allowed to interact, then the helicity can be flipped.  This leads to a redistribution of the abundances in the Dirac case, $n(\nu_{h_L}) = n(\nu_{h_R}) = n(\bar{\nu}_{h_R}) = n(\bar{\nu}_{h_L}) = n_0 / 2$, but no change in the Majorana case since the heavy neutrinos are decoupled.  We will return to this point in \sref{sec:clustering} when we discuss gravitational clustering.

\subsection{Detection of the \cnb}
\label{sec:cnb_detection}

In this section the rate of \cnb\ capture on tritium is worked out.  To best illustrate the role of helicity eigenstates, 
we start by discussing the case of the more elementary process of \n\ scattering on a neutron, and then generalize to the case of tritium.
\\

\noindent 
{\it (i) neutrino absorption on a free neutron. }  \\
Let us consider the process
\begin{align}\label{eq:nu_on_n}
	\nu_j + n \to p+e^{-}~, 
\end{align}
where the incident neutrino is taken to be in a mass eigenstate $\nu_j$, following the discussion in the previous section. For this process, the kinematics can be easily worked out in the rest frame of the neutron.  As per the discussion of \sref{sec:cnb_evolution}, the neutrino is very non-relativistic, so we can take $E_{\nu} \approx m_{\nu}$.  
After properly including the recoil of the proton, we find that the electron is ejected with a kinetic energy $K_e = E_e - m_e$, given by (see \aref{app:Kinematics})  
\begin{align}
\label{ke}
	K_{e}^{\CNB} \approx K_{\rm end} + 2 m_{\nu} \com
\end{align}
where
\begin{align}
\label{kendeq}
	K_{\rm end} 
	= \frac{ ( m_{n} - m_{e} )^2 - (m_{\nu} + m_{p})^2 }{2m_{n}} 
	= Q - \frac{m_e Q}{m_{n}} - \frac{Q^2}{2 m_{p}}
\end{align}
is the beta decay endpoint energy\footnote{
Neglecting nucleon recoil is equivalent to neglecting the last two terms in in \eref{kendeq}, and gives the more familiar result $K_{e}^{\CNB} \approx Q + 2 m_{\nu}$. This approximation is not really legitimate, however, since the size of the neglected terms exceeds the \n\ mass: \eg, for $m_\nu=0$ we get $Q^0 \approx 0.7823 \MeV$, $K_{\rm end}^0 \approx 0.7816 \MeV$, and therefore $K_{\rm end}^0 - Q^0 \approx - 0.7 \keV$.}   and $Q \equiv m_{n} - m_{p} - m_{e} - m_{\nu}$.  

We calculate the scattering amplitude for the processes in \eref{eq:nu_on_n}. 
Due to the low energies involved, we can safely work in the four-fermion  interaction approximation, and obtain  (see \aref{eq:Appendix1} for details): 
\begin{align}\label{eq:matrix_element}
	i \mathcal{M}_j = -i 
	 \frac{G_F}{\sqrt{2}} V_{ud} U^*_{ej} 
	\left[ \overline{u}_{e} \gamma^{\alpha} (1 - \gamma^5) u_{\nu_{j}} \right] 
	\left[ \overline{u}_{p} \gamma^{\beta} \left( f(0) - g(0) \gamma^{5} \right) u_{n} \right]
	\eta_{\alpha \beta} \com
\end{align}
where $u_{x}$ is the Dirac spinor for species $x$, and $V_{\rm ud} \approx 0.97425$ is an element of the Cabibbo-Kobayashi-Maskawa (CKM) matrix \cite{Beringer:2012zz}.   The element $U_{ej}$ of the PMNS matrix appears because only the electron component of each mass eigenstate can participate in the process (\ref{eq:nu_on_n}). 
The functions $f(q)$ and $g(q)$ are nuclear form factors, and in the limit of small momentum transfer they approach $f \equiv f(0) \approx 1$ and $g \equiv g(0) \approx 1.2695$ \cite{Beringer:2012zz}.

We proceed to calculate the cross section by squaring the amplitude and performing the appropriate spin sums.  
In the neutrino capture experiment under consideration, the spins of the final state electron and nucleus are not measured, and therefore we must sum over the possible final states.  
Similarly, the initial nucleus is not prepared with a definite spin, and therefore we must sum over its two possible spins.  
However, as we discussed in \sref{sec:cnb_evolution}, Dirac neutrinos are prepared in a definite spin state, they are left-helical, whereas both helicities are present if the neutrinos are Majorana.  
We will keep the calculation general for now.  
We denote the neutrino helicity by $s_{\nu}$ where $s_{\nu} = +1/2$ corresponds to right-handed helicity and $-1/2$ to left-handed.  

Having performed the spin sums as discussed above, one finds the squared matrix element to be (see \aref{eq:Appendix1} for details) 
\be \label{eq:squared_amp}
\overline{\abs{\mathcal{M}}_j^2}(s_\nu) = 8 G_F^2 |V_{ud}|^{2} |U_{ej}|^{2} m_n m_p E_e E_{\nu} \Bigl[ 
	A(s_{\nu}) (f^2 + 3 g^2 )  
	+ B(s_{\nu}) (f^2 - g^2) v_{e} \cos \theta \Bigr] \com
\ee
where $\theta$ is the angle between the \n\ and electron momenta, $\cos \theta = {\bf p}_e \cdot {\bf p}_{\nu_j} / ( \abs{{\bf p}_{e}} \abs{{\bf p}_{\nu_j}} )$, and $v_{i}$ is the velocity of the species $i$: $v_i \equiv \abs{{\bf p}_i} / E_i$.  
The spin-dependent factors are 
\begin{align}
	A(s_{\nu}) 
	&\equiv 1 - 2 s_{\nu} v_{\nu_j}
	= \begin{cases}
	1 - v_{\nu_j} & , \quad s_{\nu} = + 1/2 \qquad \text{right helical} \\
	1 + v_{\nu_j} & , \quad s_{\nu} = - 1/2 \qquad \text{left helical} \com
	\end{cases} \nonumber \\
	B(s_{\nu}) 
	&\equiv \, v_{\nu_{j}} - 2 s_{\nu} 
	\ = \begin{cases}
	v_{\nu_j} - 1  & , \quad s_{\nu} = + 1/2 \qquad \text{right helical} \\
	v_{\nu_j} + 1 & , \quad s_{\nu} = - 1/2 \qquad \text{left helical} 
	\end{cases}  \label{eq:v_cases} \per
\end{align}
If the neutrinos were relativistic, $v_{\nu_j} \simeq 1$, then we would find $A=B=0$ for right-helical neutrinos, which implies that these particles cannot be captured, and  $A = B = 2$ for left-helical neutrinos.  
This reproduces the familiar finding that  in the relativistic limit helicity and chirality coincide, and only the left-chiral neutrinos interact with the weak force.  
In the non-relativistic limit, which is relevant here, we have $A(\pm 1/2) = \mp B(\pm1/2) = 1$, indicating that both left- and right-helical neutrinos can be captured.

We calculate the differential cross section from the squared amplitude, \eref{eq:squared_amp}, in the standard way (see \aref{eq:Appendix1}), and get:
\begin{align}\label{diffcross}
	\frac{d \sigma_j (s_\nu)}{d \cos \theta} = \frac{G_F^2}{4\pi}  |V_{ud}|^{2} |U_{ej}|^{2} F(Z, E_e) \frac{m_p E_{e} p_{e}}{m_n v_{\nu_j}} \Bigl[ A(s_{\nu}) (f^2 + 3 g^2 ) + B(s_{\nu})  (f^2 - g^2) v_{e} \cos \theta \Bigr] 
\end{align} 
where $F(Z,E_e)$ is the Fermi function describing the enhancement of the cross section due to the Coulombic attraction between the outgoing electron and proton. It can be modeled as \cite{Fukugita-Yanagida}
\be\label{fermifunction}
	F(Z, E_e) = \frac{2 \pi \eta}{1- e^{-2\pi \eta}} \com
\ee
with $\eta = Z \alpha E_e/p_e$, and $Z$ being the atomic number of the daughter nucleus ($Z=1$ here); $\alpha \approx 1/137.036$ is the fine structure constant.  

Since the incoming neutrino is practically at rest, $p_{\nu} \ll p_{e}$, the kinematics allow for isotropic emission of the electron.  
Then the integral over $\theta$ is trivial, and one obtains the total capture cross section multiplied by the \n\ velocity, which is the quantity relevant for the capture rate:
\begin{align}\label{cross}
	\sigma_{j}(s_\nu) v_{\nu_j} = \frac{G_F^2}{2\pi}  |V_{ud}|^{2} |U_{ej}|^{2} F(Z, E_e) \frac{m_p}{m_n} \, E_{e} \, p_{e} \, A(s_{\nu}) (f^2 + 3 g^2 ) \per
\end{align} 
Since $A(\pm 1/2) = 1$ in the approximation $v_{\nu_j} \ll 1$, the cross section is identical for the two spin states.   
Therefore any differences in the capture rate of different spin states must arise from their abundance today, as will be seen below.
\\

\noindent 
{\it (ii) neutrino absorption on tritium. } \\
Finally, let us generalize our results to the process
\begin{align}
	\nu_j + \tensor[^3]{\rm H}{} \to \tensor[^3]{\rm He}{} + e^{-} \per
\end{align}
The calculation of the cross section runs parallel to the derivation of \eref{cross}, upon replacing $n \to \tensor[^3]{\rm H}{}$ and $p \to \tensor[^3]{\rm He}{}$.  The neutron and proton masses are replaced with the nuclear masses of the species involved: $m_n \to m_{\tensor[^3]{\rm H}{}} \approx 2808.92 \MeV$ and $m_p \to m_{\tensor[^3]{\rm He}{}} \approx 2808.39 \MeV$.  The same replacement must be done in Eqs. (\ref{ke}) and (\ref{kendeq}) to find the Q-value and the beta spectrum endpoint. Neglecting the \n\ mass, these evaluate to\footnote{
We would like to stress that one expects to find the \cnb\ signal at an energy that is displaced by $2 m_{\nu} = O(0.1 \eV)$ above the beta decay endpoint, $K_{e} = K_{\rm end}$, and that the endpoint itself is displaced by $3.4 \eV$ below the $Q$-value of the decay.  Since $3.4 \eV \gg m_{\nu}$ one should take care not to confuse the endpoint and the Q-value.  
}:  
\begin{align}\label{eq:Kend0}
	Q^{0} \approx 18.6 \keV 
	\qquad {\rm and} \qquad
	K_{\rm end}^{0} - Q^{0} \approx -3.4 \eV \per
\end{align}
Instead of the form factors, $f(q)$ and $g(q)$, one now encounters nuclear matrix elements that quantify the probability of finding a neutron in the $\tensor[^3]{\rm H}{}$, on which the neutrino can scatter, and a proton in the $\tensor[^3]{\rm He}{}$.  
This requires the replacement $f^2 \to \langle f_{F} \rangle^2 \approx 0.9987$ and $3g^2 \to (g_A/g_V)^2 \langle g_{GT} \rangle^2$ where $\langle g_{GT} \rangle^2 \approx 2.788$, $g_A \approx 1.2695$, and $g_V \approx 1$ \cite{Schiavilla:1998je}.  

After making the replacements described above, we obtain the velocity-multiplied capture cross section for mass eigenstate $j$:  
 \be\label{sigma}
	\sigma_{j}(s_\nu) v_{\nu_j} = A(s_{\nu}) \, |U_{ej}|^{2} \bar{\sigma} \com
\ee 
where
\begin{align}\label{eq:barsigma0}
	\bar{\sigma} \equiv \frac{G_F^2}{2\pi}  |V_{ud}|^{2} F(Z,E_e) \frac{m_{\tensor[^3]{\rm He}{}} }{m_{\tensor[^3]{\rm H}{}}} \, E_{e} \, p_{e} \, \Bigl( \langle f_{F} \rangle^2 + \left( g_A / g_V \right)^2 \langle g_{GT} \rangle^2 \Bigr)\simeq  3.834 \times 10^{-45} \cm^{2} \per
\end{align}
In the numerical estimate we use $E_e = m_e + K_{e}^{C\nu B}$ and \eref{kecnb}.  
Considering that for non-relativistic \ns, $A(+1/2)=A(-1/2)=1$, we obtain again that the capture cross section is the same for the left- and right-helical states, and is given by:
\begin{align}
	\sum_{j=1,2,3} \sigma_j(s_{\nu} = \pm1/2) v_{\nu_j} \Bigr|_{v_{\nu_j} \ll 1} = \bar{\sigma} \com
\end{align}
after summing over the mass eigenstates and using the unitarity of the PMNS matrix, $\sum_j | U_{ej} |^2 = 1$.  

To clarify possible confusions, it is worth noting how this result is related to other commonly encountered cross sections, namely:\\

\noindent 
{\it (i) the spin-averaged and mass summed cross section}\\
This cross section is velocity-independent, because $A(+1/2) + A(-1/2) = 2$ independent of $v_{\nu_j}$, and is:
\begin{align}
	\frac{1}{2} \sum_{s_{\nu} = \pm 1/2} \sum_{j=1,2,3} \sigma_j(s_{\nu}) v_{\nu_j} = \bar{\sigma} \com
\end{align}

\noindent
{\it (ii) the cross section to capture relativistic neutrinos} \\
This cross section vanishes for the right-helical state and for the left-helical state it is equal to twice our result:
\begin{align}\label{eq:sigma_relativistic}
	\sum_{j=1,2,3} \sigma_j(s_{\nu} = -1/2) v_{\nu_j} \Bigr|_{v_{\nu_j} = 1} = 2\bar{\sigma} =7.6 \times 10^{-45} \cm^2 \per
\end{align}
A cross section of this value has been used before in the context of \cnb\ capture on tritium in both Refs.~\cite{Cocco:2007za} and \cite{Lazauskas:2007da}, and the followup works in Refs.~\cite{Blennow:2008fh,Li:2010sn,Faessler:2011qj}. We emphasize that this is leads to an overestimate of the capture rate, and therefore it should be avoided.

Moving on, finally we can calculate the total capture rate expected in a sample of tritium with mass $M_{\rm T}$. 
In \eref{sigma} we have the capture cross section for a given neutrino mass and helicity eigenstate.  
This requires summing over the cross section for each of the six initial states ($j = 1,2,3$ and $s_{\nu} = \pm 1/2$) weighted by the appropriate flux:  
\be 
	\Gamma_{\CNB} = \sum_{j=1}^{3} \left[ \sigma_{j}(+1/2) \, v_{\nu_j} \, n_{j} (\nu_{h_R})  + \sigma_{j}(-1/2) \, v_{\nu_j} \, n_{j}(\nu_{h_L}) \right] N_{\rm T} \com
\ee
where $N_{\rm T} = M_{\rm T} / m_{\tensor[^3]{\rm H}{}}$ is the approximate number of nuclei in the sample.  
Using \eref{sigma} the capture rate can be written as 
\begin{align} \label{eq:Gamma_CNB}
	\Gamma_{\CNB} = \sum_{j=1}^{3} |U_{ej}|^2 \, \bar{\sigma} \left[ n_{j} (\nu_{h_R})  + n_{j}(\nu_{h_L}) \right]  N_{\rm T}  = \bar{\sigma} \bigl[ n(\nu_{h_R}) + n(\nu_{h_L}) \bigr] N_{\rm T} \com
\end{align}
where $\bar{\sigma}$ was given by \eref{eq:barsigma0}, and we used the fact that different \n\ mass eigenstates are equally populated \cite{Gnedin:1997vn} to perform the sum over $j$. Here $n(\nu_{h_L})$ and $n(\nu_{h_R})$ are the number densities of left- and right-helical neutrinos per degree of freedom. We have also used $A(-1/2) \approx A(+1/2) \approx 1$ in the non-relativistic limit.  

\eref{eq:Gamma_CNB} is the central result of this section.  Let us see how it applies to the cases of Dirac and Majorana neutrinos, using the results of \sref{sec:cnb_evolution}.  If the neutrinos are Dirac particles, we saw that $n(\nu_{h_L}) = n_0$ and $n(\nu_{h_R}) = 0$, and the capture rate becomes
\be\label{Drate}
	\Gamma_{\CNB}^{\rm D} = \bar{\sigma} n_{0} N_{\rm T}  \per
\ee
Alternatively, for the Majorana case we found $n(\nu_{h_L}) =n(\nu_{h_R})= n_0$, and the capture rate becomes
\be\label{Mrate}
	\Gamma_{\CNB}^{\rm M} = 2 \bar{\sigma} n_{0} N_{\rm T} \per
\ee
That is, the capture rate in the Majorana case is twice that in the Dirac case:
\begin{align}\label{eq:M_is_2D}
	\Gamma_{\CNB}^{\rm M} = 2 \, \Gamma_{\CNB}^{\rm D} \per
\end{align}
The relative factor of $2$ is a central result of our paper.  It can be understood as follows.  In the Dirac case, we found that the \cnb\ consists of only left-helical neutrinos and right-helical anti-neutrinos.  If these neutrinos were in the relativistic limit, where helicity and chirality coincide, only the left-helical states could interact weakly.  The right-helical states would be sterile, and only half of the background neutrinos would be available for capture.  Since the \cnb\ is non-relativistic, both the left- and right-helical states contain some left-chiral component, and therefore they both interact.  The right-helical anti-neutrinos cannot be captured because the process $\bar{\nu} + p \to n + e^{+}$ is kinematically forbidden: it requires $E_{\nu} > (m_n + m_e - m_p) \approx 2 \MeV$ in the proton rest frame, but the \cnb\ neutrinos only carry $E_{\nu} \approx m_{\nu} \lesssim \eV$ (similarly for the tritium).  Thus in the Dirac case, only half of the \cnb\ abundance is available for capture.  On the other hand, for the Majorana case one does not distinguish neutrinos and anti-neutrinos; instead we find that the \cnb\ consists of left-helical neutrinos and right-helical neutrinos, which both interact weakly and therefore are available for capture.


\section{Detection prospects at a PTOLEMY-like experiment}
\label{sec:ptolemy}

Let us now turn to the phenomenology of a tritium-based experiment. Considering a target mass of 100 g, as is proposed for PTOLEMY \cite{Betts:2013uya}, Eqs.~(\ref{Drate}) and (\ref{Mrate}) evaluate to 
\be \label{eq:signal_rates}
	\Gamma_{\CNB}^{\rm M} \approx 8.12~ {\rm yr}^{-1} 
	\qquad {\rm and} \qquad  
	\Gamma_{\CNB}^{\rm D} \approx 4.06 ~ {\rm yr}^{-1} 
\ee
for the Dirac and Majorana neutrino cases, respectively. These rates are limited only by the sample size, since they are independent of the \n\ mass (as long as the \ns\ are non-relativistic), and the \cnb\ neutrino flux is fixed (in absence of exotica).

One of the main challenges for a neutrino capture experiment is the energy resolution. The resolution of a detector quantifies the smallest separation at which two spectral features (\eg, two peaks) can be distinguished.  For instance, two Gaussian curves centered at $E_1$ and $E_2$, having equal amplitude, and having equal standard deviation $\sigma$ can be distinguished provided that $|E_1-E_2| \gtrsim \Delta$ where 
\be
\Delta = \sqrt{8 \ln 2} \sigma \approx 2.35 \sigma
\label{fwhm}
\ee
is the full width at half maximum (FWHM) of the Gaussian \cite{WilliamLeoBook}. The FWHM is conventionally taken to be the detector resolution.  Applied to our case, this argument means that the spectral excess due to the \cnb\ can be resolved if its separation from the beta endpoint exceeds the resolution: $\Delta K_e = 2 m_{\nu} \gta \Delta$. 
  
PTOLEMY is expected to achieve an energy resolution of $\Delta = 0.15 \eV$ \cite{Betts:2013uya}, just enough to  probe the upper end of the \n\ mass spectrum, where the three masses $m_j$ are degenerate or quasi degenerate: $|m_i - m_j|\ll m_j$, \ie, $m_1 \simeq m_2 \simeq m_3 = m_\nu$.  In this situation, the mass splittings can not be resolved by the detector, and the signature of the \cnb\ reduces to a single excess corresponding to the effective mass $m_\nu$.  Most of the discussion from here on will refer to this case.   A brief discussion on possibly resolving the individual masses is given in \sref{sub:hierarchy}.

Tritium beta decay is the best known and likely the main source of background\footnote{See Ref.~\cite{Betts:2013uya} for a discussion of additional backgrounds.} for the \cnb\ neutrino capture events. The effect of the finite energy resolution is that the most energetic electrons from beta decay might have {\it measured} energy that extends beyond the endpoint $K_{\rm end}$, into the region where the signal is expected.  

 To estimate the rate of such events, consider first the beta decay spectrum  \cite{Masood:2007rc}:
\begin{align}\label{gammabeta}
\frac{d\Gamma_\beta}{dE_e} = \sum_{j=1}^{3} |U_{ej}|^{2} \frac{\bar{\sigma}}{\pi^2} H(E_e, m_{\nu_{j}}) N_{\rm T} \com
\end{align}
where 
\begin{align}
	H(E_e, m_{\nu_{j}}) \equiv \frac{1 - m_e^2 / (E_e m_{\tensor[^3]{\rm H}{}}) }{(1 - 2 E_e / m_{\tensor[^3]{\rm H}{}} + m_e^2 / m_{\tensor[^3]{\rm H}{}}^2 )^2} \sqrt{y \left( y + \frac{2m_{\nu_{j}} m_{\tensor[^3]{\rm He}{}}}{m_{\tensor[^3]{\rm H}{}}} \right) } \left[ y + \frac{m_{\nu_{j}}}{m_{\tensor[^3]{\rm H}{}}} \left(m_{\tensor[^3]{\rm He}{}} + m_{\nu_{j}} \right) \right] \com
\end{align}
and $y = m_e + K_{\rm end} - E_e$ and the other variables are as in \sref{sec:cnb_detection}.  
  
After integrating over energy, the total tritium beta decay rate is found to be
\begin{align}
	\Gamma_{\beta} = \int_{m_e}^{m_e + K_{\rm end}} dE_e \, \frac{d\Gamma_{\beta}}{dE_e} \approx  10^{24} \left( \frac{M_T}{100~{\rm g}}\right) \yr^{-1} \per
	\end{align}
Comparing with the signal rate in \eref{eq:signal_rates}, it appears immediately that even an extremely small contamination of beta decay events in the signal region can represent a serious challenge for \cnb\ detection.

\begin{figure}[t]
\begin{center}
\includegraphics[width=0.48\textwidth]{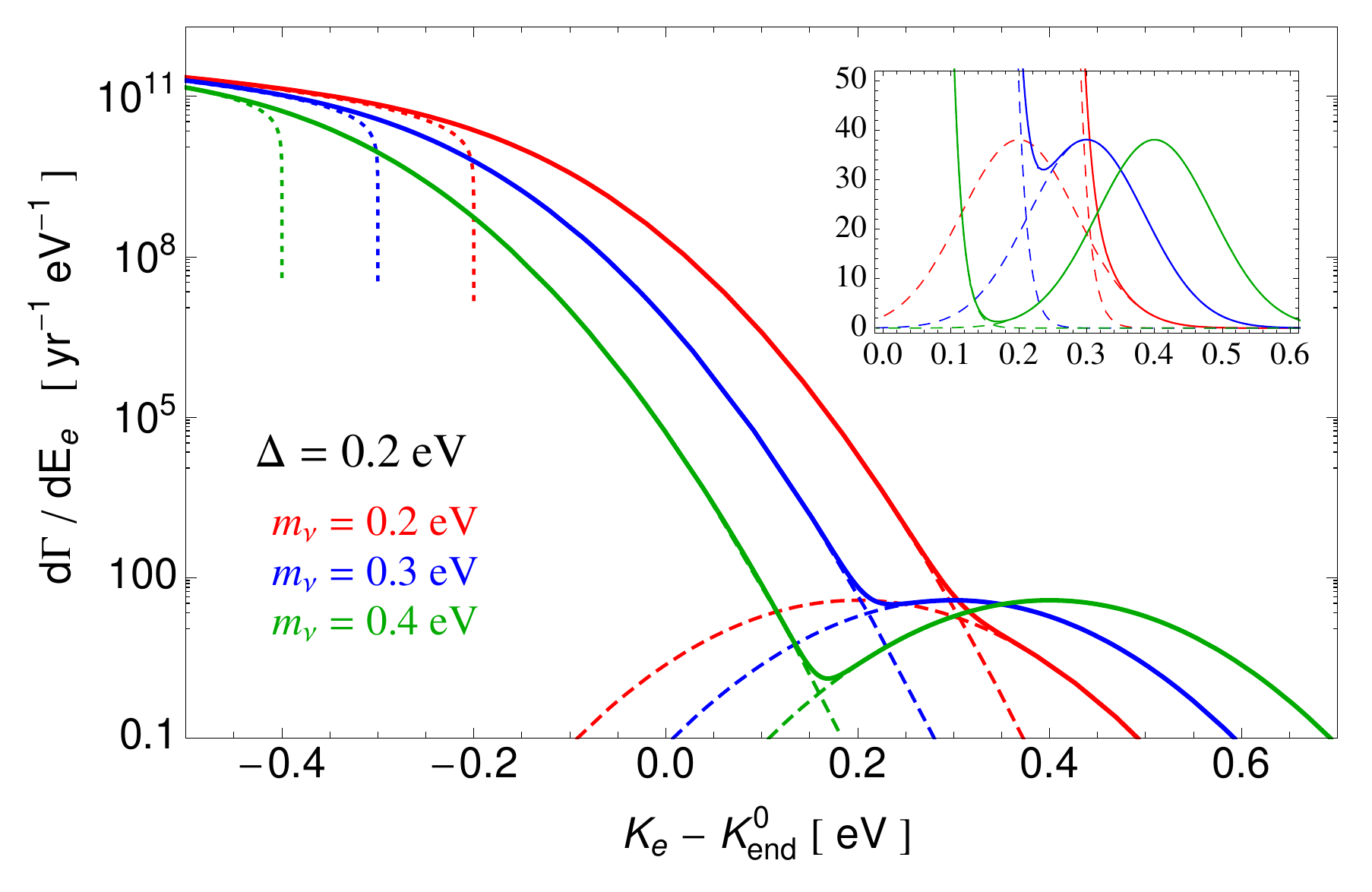} \hfill
\includegraphics[width=0.48\textwidth]{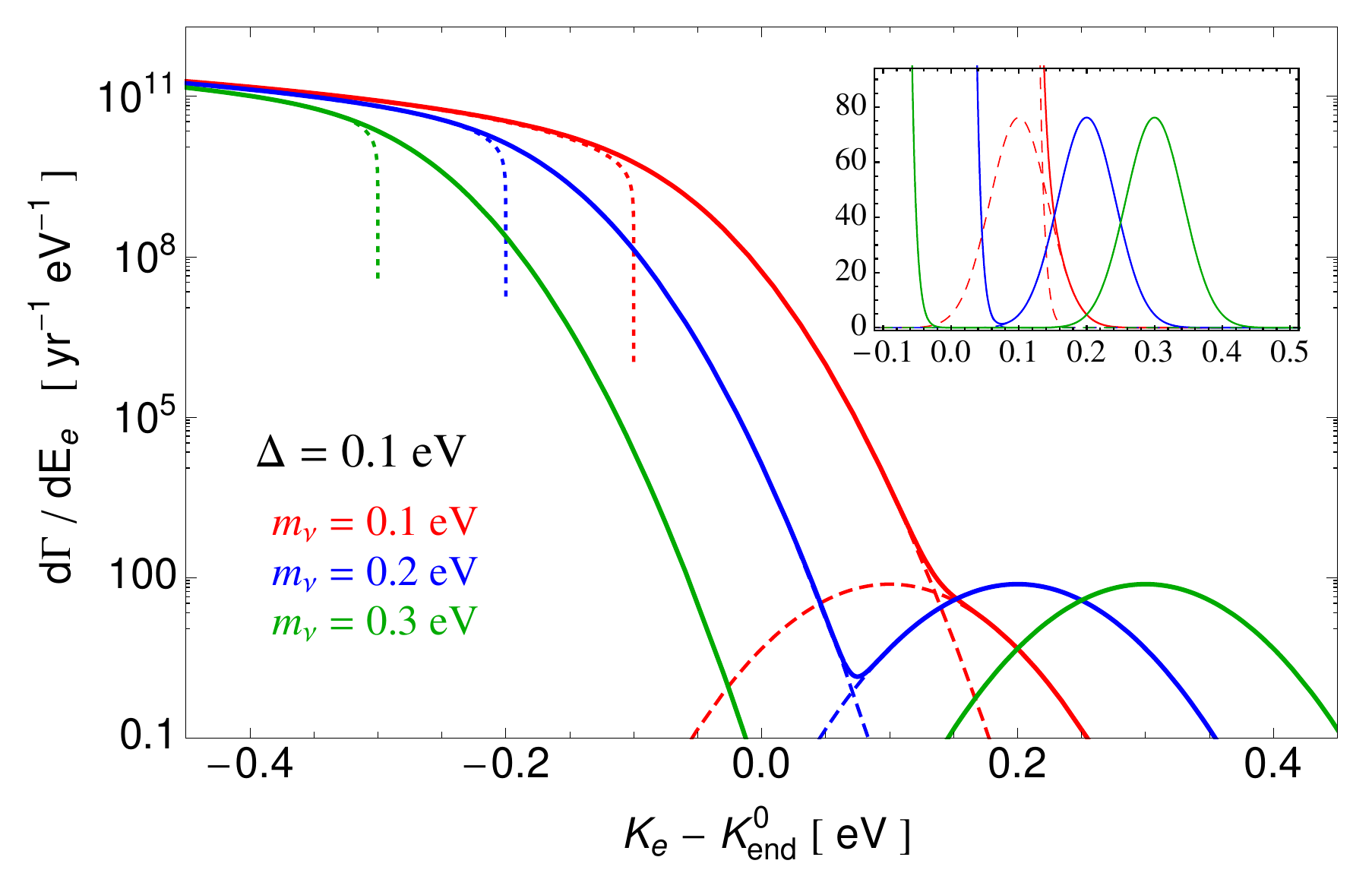} 
\caption{\label{fig:spectrum}
Solid lines: the expected spectrum of electrons in terms of observed energy, obtained from Eqs.~(\ref{smearedgammacap}) and (\ref{smearedgammacapbeta}), for detector resolution (FWHM) $\Delta$ and neutrino mass $m_{\nu}$.  The dashed lines give the two contributions (signal and background) separately.  The dotted lines show the spectrum of beta decay electrons for the ideal case of perfect energy resolution, $\Delta \simeq 0$.   The zero of the horizontal axis coincides with beta decay endpoint (for perfect resolution) for massless neutrinos. 
}
\end{center}
\end{figure}

To calculate the number of background events, we model the {\it observed} spectrum by convolving the beta decay and \cnb\ event ``true"  spectra  with a Gaussian envelope of FWHM $\Delta$ [\eref{fwhm}]:
\begin{align}
\label{smearedgammacap}
	\frac{d \tilde{\Gamma}_{\CNB}^{\rm M}}{dE_e} 
	&= \frac{1}{\sqrt{2 \pi} \, \sigma} \int_{-\infty}^{\infty} dE_{e}^\prime~ \Gamma_{\CNB}^{\rm M} (E_e^\prime) ~\delta[E_e^\prime - (E_{\rm end} + 2m_{\nu})]~ \exp \Bigl[ - \frac{(E_e^{\prime} - E_e)^2}{2 \sigma^2} \Bigr] \\
	\label{smearedgammacapbeta}
	\frac{d \tilde{\Gamma}_{\beta}}{dE_e} 
	&= \frac{1}{\sqrt{2 \pi} \, \sigma} \int_{-\infty}^{\infty} dE_{e}^\prime~ \frac{d \Gamma_{\beta}}{d E_e}(E_e^\prime) \, \exp \Bigl[ - \frac{(E_e^{\prime} - E_e)^2}{2 \sigma^2} \Bigr] \per
\end{align}
In \fref{fig:spectrum} we show the smoothed spectra and their sum for various different combinations of detector resolution and neutrino mass.  For $\Delta \approx m_{\nu}$, the smoothed beta decay spectrum extends well beyond the endpoint energy at $K_{e} - K_{\rm end}^{0} \approx - m_{\nu}$ and contaminates the neutrino capture signal region at $K_{e} - K_{\rm end}^{0} \approx + m_{\nu}$.  

\begin{figure}[t]
\begin{center}
\includegraphics[width=0.65\textwidth]{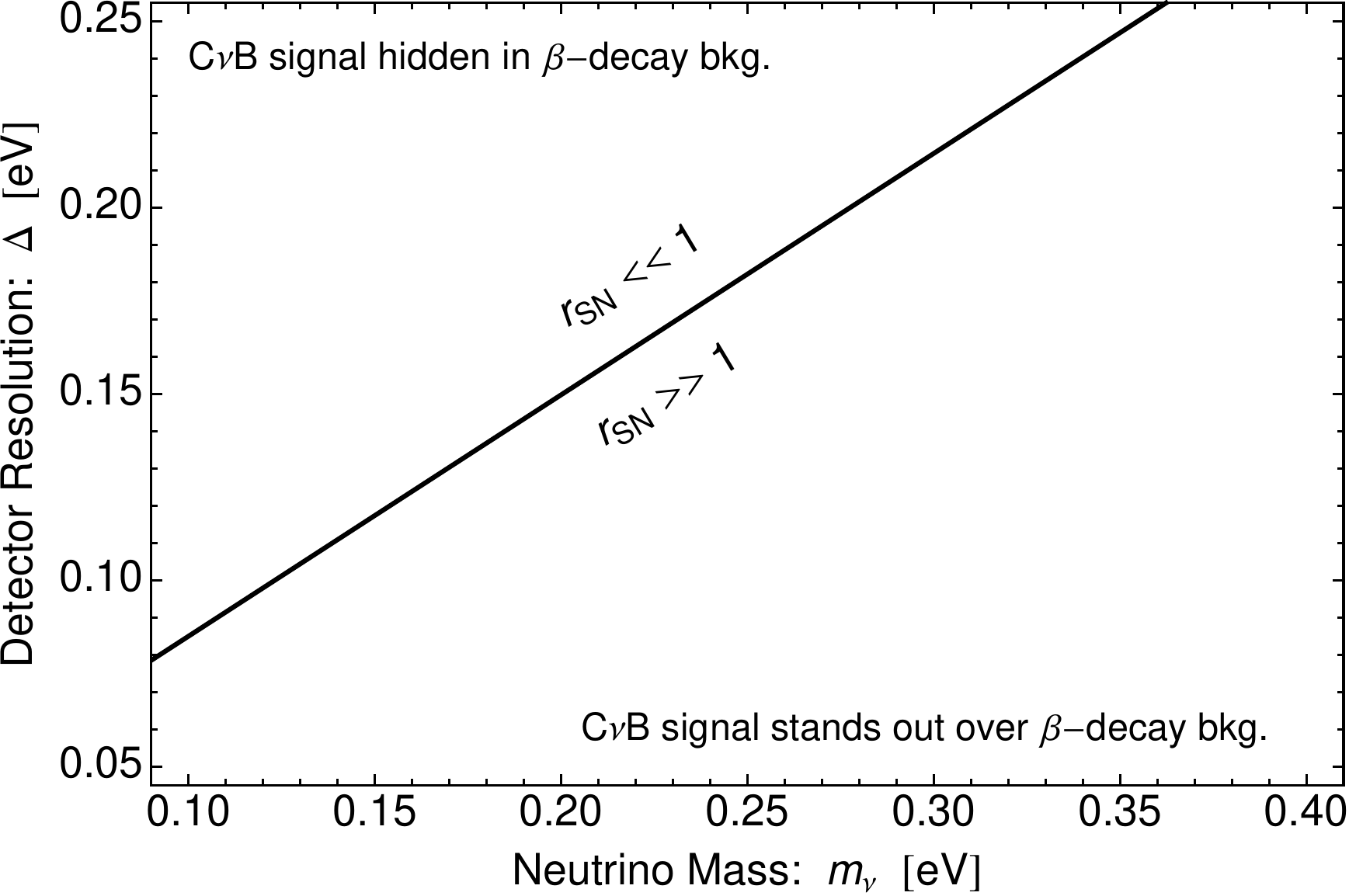} 
\caption{
\label{fig:snr}
Contour plot of signal to noise ratio, $\rSN =1$ [\eref{snr}], for a range of detector resolutions, $\Delta$, and neutrino masses, $m_\nu$, for Majorana \ns\ with a degenerate neutrino mass spectrum.  In the region below the $\rSN =1$ line, the \cnb\ signal stands out over the beta decay background, and in the region above this line, the background events dominate. For Majorana \ns, $\rSN\gtrsim 1$ corresponds to $\Delta \lesssim 0.7 m_\nu $.
}
\end{center}
\end{figure}

To estimate the potential to distinguish the signal  from the background, we calculate the  signal-to-noise ratio. 
Following \cite{Cocco:2007za}, the calculation is done for an (observed) energy bin  of width $\Delta$ that is centered on the neutrino capture signal peak. In this bin, the signal and background event rates are: 
\begin{align}\label{gammacnbdelta}
	\tilde \Gamma_{\CNB}^{\rm M} (\Delta) 
	& = \int_{E_{e}^{\CNB} - \Delta/2}^{E_{e}^{\CNB} + \Delta/2} dE_{e} \,  \frac{d \tilde{\Gamma}_{\CNB}}{dE_e}(E_e) \com \\
\label{gammabetadelta}
	\tilde \Gamma_\beta (\Delta) 
	& = \int_{E_{e}^{\CNB} - \Delta/2}^{E_{e}^{\CNB}+ \Delta/2} dE_{e} \,  \frac{d \tilde{\Gamma}_{\beta}}{dE_e}(E_e) \com
\end{align}
respectively, where $E_{e}^{\CNB} \equiv K_{e}^{\CNB}+ m_e + 2 m_{\nu}$, and their ratio is:  
\be\label{snr}
	\rSN = \frac{\tilde{\Gamma}_{\CNB}^{\rm M} (\Delta)} {\tilde{\Gamma}_\beta (\Delta)} \per
\ee
In \fref{fig:snr}, contour plot of $\rSN =1$ is shown for a range of detector resolutions and neutrino masses.  Successful detection of the \cnb\ signal is impossible if $\rSN \ll 1$, and it is very likely if $\rSN \gg 1$.  For a given $\Delta$, the signal-to-noise ratio is a rapidly rising function of the neutrino mass (and therefore of the width of the gap in energy between the \cnb\ signal and the beta spectrum endpoint), because the endpoint electrons are exponentially suppressed in the tail of the Gaussian.  As a rule of thumb, for Majorana \ns\ we find that 
\begin{align}\label{eq:Delta_bound}
	\rSN\gtrsim 1
	\qquad {\rm for} \qquad 
	\Delta \lesssim 0.7 m_\nu \per
\end{align}
This condition is only slightly different for Dirac \ns, although the signal rate itself is lower by a factor of 2 [\eref{eq:signal_rates}].  

This conclusion on the signal-to-noise ratio differs slightly from that in the similar analysis of \rref{Cocco:2007za}. The difference is due to two aspects: (i) here  $\rSN$ is obtained by numerically evaluating Eqs.~(\ref{gammacnbdelta}) and (\ref{gammabetadelta}). Instead, in \rref{Cocco:2007za}  the convolution integral is approximated by a factorized form for the beta decay background, which tends to underestimate $\rSN$, and the \cnb\ signal was not convolved with a Gaussian, which tends to overestimate the signal.   (ii) here  $\Delta$, is identified with the Gaussian FWHM (under the advice of the PTOLEMY collaboration, \cite{TullyPC}), and not with the Gaussian standard deviation $\sigma$ as in \rref{Cocco:2007za}. In terms of $\sigma$, our condition reads $m_{\nu} \gtrsim 1.4 (\sqrt{8 \ln 2} \sigma) \approx 3.3 \sigma$, which is compatible with \rref{Cocco:2007za}. 

In \tref{tabactive} we consider various values for the detector resolution and neutrino masses, and we show the expected signal event rates and signal-to-noise ratios for the Dirac and Majorana cases.  
We also show the effect of neutrino clustering; see \sref{sec:clustering} below.  
If $\rSN$ is large then the systematic error arising from the beta decay endpoint is negligible, and (in the absence of other systematic errors) the limiting factor is statistics.  
If $N$ events are detected, then the counting error is expected to go like $\sqrt{N}$, and the statistical significance can be estimated as $N / \sqrt{N} = \sqrt{N}$.  
A $3\sigma$ detection requires $N \approx 9$ events in the signal region, and a $5\sigma$ detection requires $N \approx 25$ events.  
With an event rate of $\Gamma_{\CNB}^{\rm M} \approx 8 \, {\rm yr}^{-1}$ these significances would require approximately $1$ and $3$ years of data taking, respectively.  
\begin{table}[h]
\begin{center}
\begin{tabular}{|c|c|c|c|c|c|c|c|}
\hline
$\Delta~{\rm (eV)}$ & $m_{\nu}~{\rm (eV)}$ & $\Gamma_{\CNB}^{\rm D}~{\rm (yr^{-1})}$ & $\rSN^{\rm D}$ & $\Gamma_{\CNB}^{\rm M}~{\rm (yr^{-1})}$ & $\rSN^{\rm M}$ & $f_{\rm c}^{\rm NFW}$ & $f_{\rm c}^{\rm MW}$ \\
\hline
0.10 & 0.15 & \multirow{4}{*}{4.1 (3.1)} & 37 & \multirow{4}{*}{8.1 (6.2)} & 74 & 1.4 & 1.6 \\
0.20 & 0.30 &  & 4.6 &  & 9.2 & 3.1 & 4.4 \\
0.30 & 0.45 &  & 1.4 &  & 2.8 & 6.4 & 10 \\
0.40 & 0.60 &  & 0.6 &  & 1.2 & 12 & 20 \\
\hline
\end{tabular}
\end{center}
\caption{The signal to noise ratio $\rSN$ [\eref{snr}] the neutrino capture rate, and the enhancement factor due to the gravitational clustering [\eref{fclus}] for various values of the detector resolution, $\Delta$, and of the neutrino mass $m_\nu$.  M and D stand for Majorana and Dirac \ns, respectively.  The rates in parentheses refer to the neutrino capture event rates in the bin of width $\Delta$ centered at $K_e^{\CNB} = K_{\rm end} + 2 m_{\nu}$ [see \eref{gammacnbdelta}].  
\label{tabactive}
} 
\end{table}


\section{Detection prospects for varying neutrino properties}
\label{sec:outcome}

So far, we have discussed the simplest, ``base" case of capture of \ns\ with a single mass and known density given by the cosmological prediction, $n_0 = 56~{\rm cm^{-3}}$ per species. Here we elaborate further, and give a more detailed discussion of the phenomenology that is expected depending on the \n\ properties.  Specifically, we discuss the distinction between Majorana and Dirac neutrinos, the correction to the rate due to \n\ clustering, and the effect of the \n\ mass hierarchy.

\subsection{Majorana vs. Dirac neutrinos}

When neutrinos are non-relativistic, the distinction between the Dirac and Majorana character becomes pronounced.  
It is critical to recognize that the \cnb\ represents the only known source of non-relativistic neutrinos in the universe.  
As we saw in \sref{sec:cnb_detection}, the Dirac or Majorana character of the \cnb\ neutrinos has a significant effect on \cnb\ neutrino capture:  the capture rate for Majorana neutrinos is double that of Dirac neutrinos [see \eref{eq:M_is_2D}].  
The factor of two difference can be understood as follows.  
For the Dirac case, only the left-helical neutrinos are available for capture since the right-helical neutrino population is absent from the \cnb, and the anti-neutrinos cannot be captured.  
For the Majorana case, the \cnb\ contains both left- and right-helical neutrinos, and the capture rate is doubled.  
In \tref{tabactive} we compare the signal rates for Dirac and Majorana neutrinos, as well as the corresponding signal to noise ratios, $\rSN$.  

\subsection{Clustering and annual modulation}\label{sec:clustering}

Like all massive particles, \ns\ should cluster in the gravitational potential wells of galaxies and clusters of galaxies.  Due to clustering, the local number density, $n_{\nu}^{c}$, is larger than the unclustered case, $n_0$, and the capture rate should therefore be enhanced by a factor
\be\label{fclus}
f_{\rm c} = \frac{n_{\nu}^{\rm c}}{n_0} \per
\ee
The calculation of $f_{\rm c}$ requires solving the Boltzmann equation for the cosmic evolution of a system consisting of both cold dark matter and neutrinos, where they are treated as warm dark matter. A variety of approaches, based on different approximations and numerical techniques, have been presented \cite{Singh:2002de, Ringwald:2004np}. We show the results of \rref{Ringwald:2004np} in the last two columns of \tref{tabactive}. There, $f_{\rm c}$ is given for two different models of the dark matter halo of our galaxy, the so called Milky Way model \cite{Klypin:2001xu} and the Navarro-Frenk-White profile \cite{Navarro:1996gj}. For masses of the order of $m_\nu \sim 0.1$ eV, the effect of clustering should be at the level of few tens of per cent, comparable to the $1\sigma$ statistical error expected at PTOLEMY in a few years or running (see \sref{sec:ptolemy}). Therefore, the experiment may not be able to measure the local value of $f_{\rm c}$, but at least it will place a first stringent constraint on it.  If the effect of clustering is indeed modest, it may be subdominant to the factor of 2 difference expected between Dirac and Majorana \ns, which could still be distinguished.  

An additional consequence of clustering is the mixing of neutrino helicities \cite{Duda:2001hd}.  
As a gravitationally bound -- but otherwise non-interacting -- neutrino orbits around the halo, its momentum changes direction and magnitude, but its spin remains fixed.  This causes helicity to change, so that a population of neutrinos initially prepared in a given helicity state (\eg, 100\% initially right-helical) will in time grow a component of the opposite helicity, and ultimately reach an equilibrium where the right-helical and left-helical states are equally populated.  We saw in \sref{sec:cnb_evolution} that the cosmological population of Dirac neutrinos (anti-neutrinos) consists of 100\% left-helical (right-helical) states [\eref{eq:n_Dirac_today}]. Assuming complete clustering (\ie, all the \ns\ available  for capture are bound gravitationally to the halo), the populations will equilibrate:  $n(\nu_{h_L}) = n(\bar{\nu}_{h_R}) = n(\nu_{h_R}) = n(\bar{\nu}_{h_L}) = n_0 / 2$. Majorana, neutrinos on the other hand are already equilibrated initially [\eref{eq:n_Maj_today}] and clustering will simply conserve the equilibrium:  $n(\nu_{h_L}) = n(\nu_{h_R}) = n_0$.  
After repeating the argument in \sref{sec:cnb_detection}, one finds that even with complete clustering the Majorana capture rate is still double that of the Dirac neutrinos. This is because for clustered Dirac \ns, the new population of right-helical states, $n(\nu_{h_R})$, compensates for the loss of the left-helical ones in \eref{eq:Gamma_CNB}.

Finally, let us consider the possibility that the \cnb\ signal rate could exhibit an annual modulation, similar to the one predicted for dark matter direct detection. This modulation could be due to the fact that if neutrinos are substantially clustered, then their velocity distribution relative to Earth is not isotropic and static, as it is usually assumed.  The modulation should then follow the relative velocity of the Earth's motion with respect to the galactic disk\footnote{
Clustering also produces a modified momentum distribution compared to unclustered \ns, specifically, for strong clustering the average momentum will be higher than that of the Femi-Dirac prediction \cite{Ringwald:2004np}.  
Additionally, the momentum distribution in the rest frame of the Earth will depend on the Earth's motion relative to the galactic plane. 
As long as the \ns\ are non-relativistic, however, changes in the \n\ momentum distribution do not affect the capture rate.  
}. 

In fact, the answer to the question of modulation is negative \cite{Safdi:2014xx}.  
As we saw in \eref{eq:Gamma_CNB}, the capture rate depends on the product of number density, cross section and \n\ velocity, $v_{\nu}$. Since neutrino capture is an exothermic process, \ie, some of the nuclear binding energy is liberated, the cross section scales as $\sigma \propto 1 / v_\nu$ \cite{Landau-Lifshitz,Lazauskas:2007da}. Since the velocity cancels in \eref{eq:Gamma_CNB}, the rate is insensitive to the neutrino velocity, and thus, there should be {\it no annual modulation} of the signal. This is different from DM direct detection, which is an elastic scattering process, with $\Gamma \propto v$. In contrast with DM, then, for \cnb\ detection the astrophysical uncertainties on the velocity profile are not an issue. In this sense, \cnb\ detection is cleaner than DM detection.  
If an annual modulation {\it does} appear at a \cnb\ detector, its origin would have to be traced elsewhere.  
For instance, an $O(0.1-1\%)$ modulation may arise from the gravitational focusing from the Sun \cite{Safdi:2014xx}, even if the neutrinos are not clustered on the scale of the Milky Way.  

\subsection{The hierarchical mass spectrum}
\label{sub:hierarchy}

Let us now consider the mass differences between the different \n\ states. 
From the observation of oscillations, the degeneracy splitting is measured to be \cite{Beringer:2012zz}: 
\begin{align}\label{eq:mass_splits}
	\Delta m_{21}^2 \approx (8.66 \meV)^2 
	\qquad {\rm and} \qquad
	\abs{ \Delta m_{32}^2 } \approx \abs{ \Delta m_{31}^2 } \approx (48 \meV)^2 \per
\end{align}
The sign of $\Delta m_{31}^2$ is yet unknown, allowing   for two possible mass hierarchies (or ``orderings"):
\begin{align}
	\text{normal hierarchy (NH):}& \quad \Delta m_{31}^2 > 0 \quad m_1 < m_2 < m_3 \\
	\text{inverted hierarchy (IH):}& \quad \Delta m_{31}^2 < 0 \quad m_3 < m_1 < m_2 \per
\end{align}
In the coming years, long baseline experiments hope to distinguish these two scenarios \cite{Ge:2012wj}.  

If the masses $m_j$ are comparable with the largest splitting, $m_j \sim \sqrt{ \abs{ \Delta m_{31}^2 } } \approx 0.05 \eV$, the degenerate, single-mass, approximation used so far becomes inadequate.  This is likely to be the case: indeed, if the stringent cosmological bound on the masses, \eref{cosmobound}, is saturated then the spectrum can only be marginally degenerate, $m_{\nu_j} \approx 0.07 \eV$.  In the hierarchical regime, \cnb\ detection will not be possible without a significant improvement in the detector resolution.  Nevertheless, we feel that it is illustrative to discuss how the signal qualitatively changes in this case.  A detailed discussion is also given in  Refs.~\cite{Blennow:2008fh,Li:2010sn}. 

\begin{figure}[t]
\begin{center}
\includegraphics[width=0.48\textwidth]{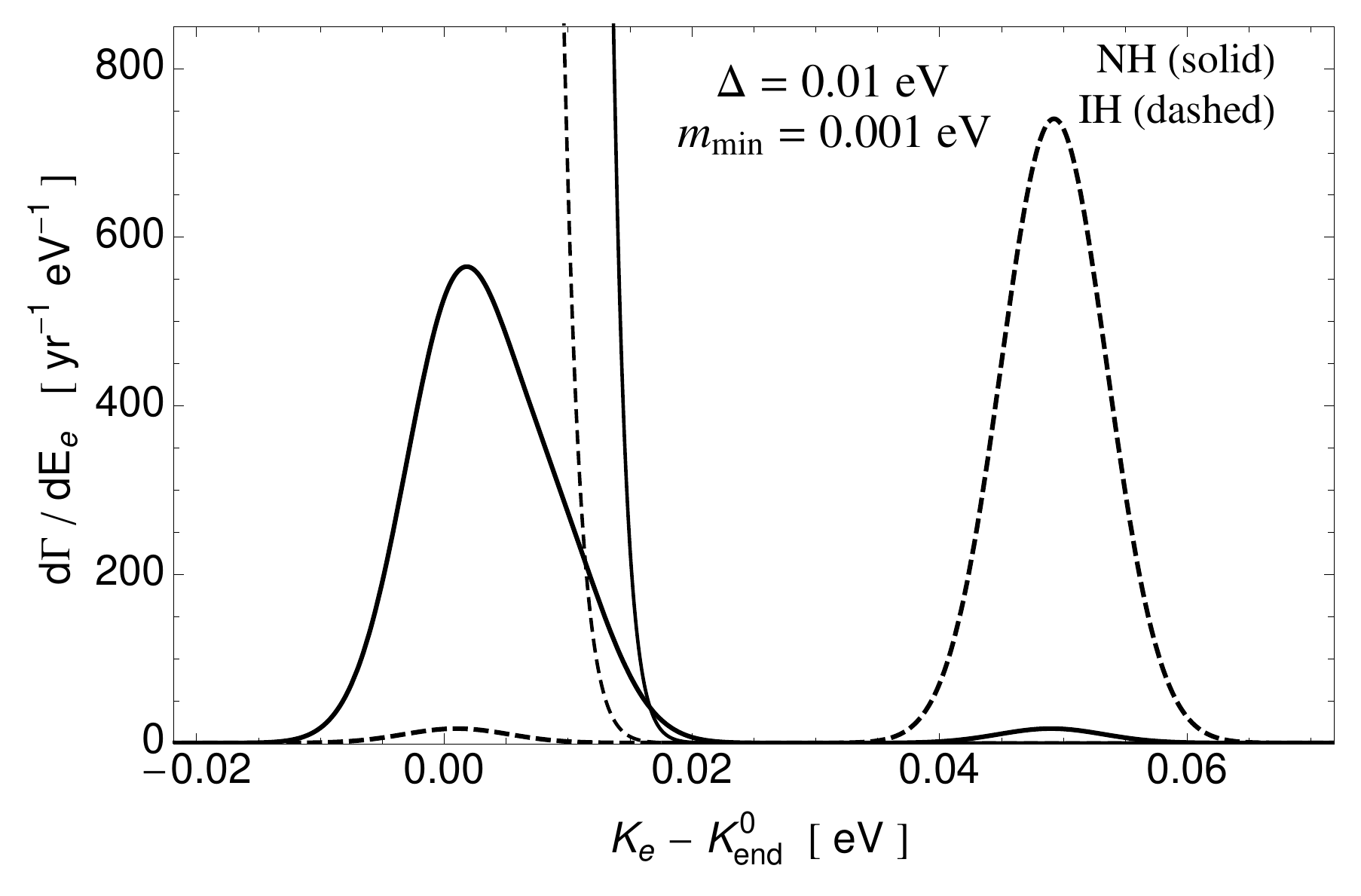} \hfill
\includegraphics[width=0.48\textwidth]{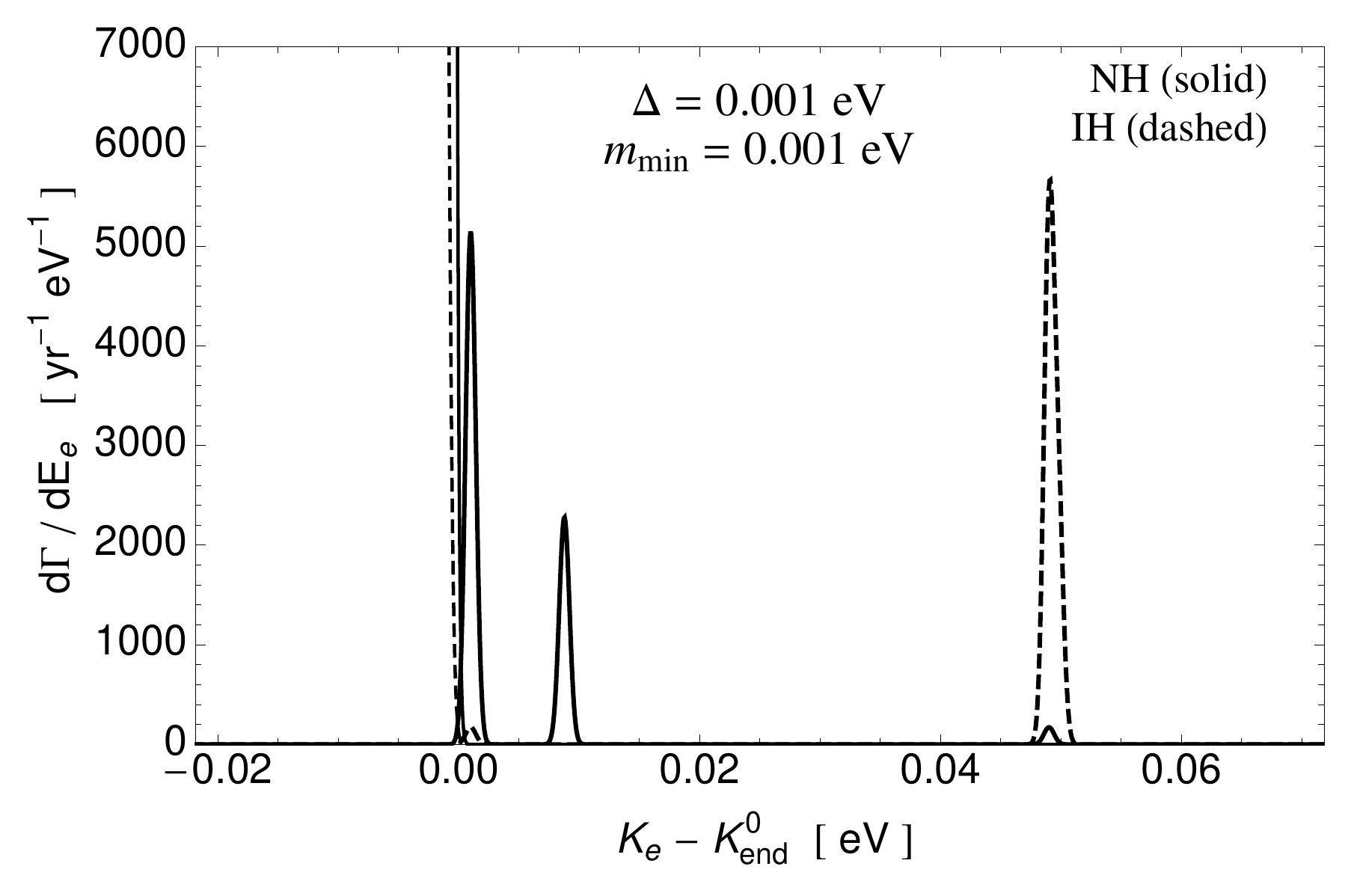} 
\caption{
\label{fig:hierarchy}
The \cnb\ signal at an ultra-high-resolution detector. Each panel shows both hierarchies (NH and IH), with lightest neutrino being almost massless, $m_{\rm min} \approx 1 \meV$.  
The Gaussian peaks are the \cnb\ signal and the sloped lines are the beta decay background.  The detector resolution is $\Delta = 0.01 \eV$ in the left panel and $\Delta = 0.001 \eV$ in the right panel.
}
\end{center}
\end{figure}

For a detector with an arbitrarily good energy resolution, $\Delta \ll m_{\nu}$, each mass eigenstate $\nu_j$ would make a distinguishable contribution to the \cnb\ capture and to the beta decay spectrum as well.  
The beta decay spectrum would be the sum of three spectra, and its endpoint would be determined by the lightest \n\ mass, $m_{\rm min} ={\rm min}[m_j]$  ($m_{\rm min}=m_1$ for NH, $m_{\rm min}=m_3$ for IH):  $K_{\rm end}=K^0_{\rm end}-m_{\rm min}$.  For the \cnb\ capture signal, each state $\nu_j$ would produce a distinct line at an electron kinetic energy of $K_{e \, j} =K^0_{\rm end}+m_{\nu_j}$, or, equivalently:
\begin{align}
	K_{e \, j}^{\CNB} = K_{\rm end} + m_{\rm min} + m_{\nu_j} \com
	\label{kej}
\end{align}
which recovers  \eref{ke} in the degenerate regime.   The total signal rate is still given by \eref{eq:Gamma_CNB}, but the three terms of the sum will appear as three separate excesses in the energy spectrum, each with weight $|U_{ej}|^2$, where \cite{Beringer:2012zz}: 
\begin{align}
	|U_{e1}|^2 \simeq 0.68 
	\qquad , \qquad 
	|U_{e2} |^2\simeq 0.30
	\qquad , \quad {\rm and} \qquad 
	|U_{e3} |^2\simeq 0.02 \per
\end{align}
Therefore, the signal is the strongest for $\nu_1$, weaker for $\nu_2$, and the weakest for $\nu_3$, as shown in \fref{fig:hierarchy}.  From the figure one can clearly see that, if we consider the effect of finite detector resolution, the \cnb\ detection is easier in the IH  case than for NH.  Indeed,  the IH case, $\nu_1$ and $\nu_2$ have the largest separation from the beta decay endpoint, and they have the strongest signal, making them easier to distinguish from the background.  In the NH case, $\nu_3$ has the largest separation, but it has the weakest signal.  Note that the intensity of the  beta decay background also differs between the IH and NH cases.  For the IH case, the endpoint is determined by $\nu_3$, which however contributes only proportionally to $|U_{e3}|^2$, hence the lower background rate. Instead, for the NH case the suppression of the beta spectrum near the endpoint is only $|U_{e1}|^2$ [see \eref{gammabeta}], corresponding to a higher background.  

In \fref{fig:hierarchy}, two values of $\Delta $ are considered.  For  $\Delta = 0.01 \eV$, in the NH case the signal is lost behind the background, but in the IH case the signal is clearly seen.  The  $\nu_2$ and $\nu_1$ eigenstates appear as a single peak, because the resolution is insufficient to resolve the small mass gap between them: $\sqrt{\Delta m_{21}^2} \approx 8.66 \meV < 0.015~{\rm eV}$. For an even more ambitious resolution, $\Delta = 0.001 \eV$, and NH, we can see the signal, and resolve both the $\nu_2$ and $\nu_3$ eigenstates.  For IH, the signal is still visible, but the $\nu_2$ and $\nu_1$ eigenstates are still not resolved. 


\section{Probing sterile neutrinos}
\label{sec:sterile}

\subsection{eV-scale sterile neutrinos}

In addition to the three known flavor eigenstates of active neutrinos, there might exist other states that are inert, or ``sterile" with respect to the Standard Model gauge interactions.  Here we discuss sterile states that mix with the active states, and share their same helicity, so that they can be produced via active-sterile oscillations. Within this scenario, the most interesting case is that of a sterile neutrino, $\nu_s$, and its corresponding mass eigenstate, $\nu_4$, with mass at the eV scale, $m_4 \sim 1 \eV$.  This additional sterile neutrino state is the favored interpretation of the anomalous excess of $\nue$ and $\barnue$ observed in $\numu$ and $\barnumu$ beams at LSND \cite{Aguilar:2001ty,Athanassopoulos:1995iw} and MiniBooNE \cite{AguilarArevalo:2008rc}.  
It is also a possible explanation of the flux deficits observed in reactor neutrinos \cite{Mention:2011rk,Mueller:2011nm,Huber:2011wv} and at solar neutrino calibration tests using gallium \cite{Anselmann:1994ar,Hampel:1997fc,Abdurashitov:1998ne}.

In presence of a fourth state, flavor mixing is described by a $4\times4$ matrix, with the elements $U_{\alpha 4}$ ($\alpha=e,\mu,\tau,s$) describing the flavor composition of $\nu_4$. The LSND / MiniBooNE experiments favor \cite{Aguilar-Arevalo:2013pmq} 
\be
	\sin^2 2 \theta = 4 |U_{e4}|^2 |U_{\mu4}|^2  \sim (1- 10 ) \cdot 10^{-3}
	\qquad {\rm and} \qquad
	\Delta m^2_{41} = m^2_4 - m^2_1 \sim (0.1 - 10) \, {\rm eV^2}~,
\label{nusparam}
\ee
while global fits of all the anomalies favor the ``democratic" value \cite{Giunti:2013aea}
\begin{align} \label{eq:Ue4}
	|U_{\mu 4}|^2 \sim |U_{e4}|^2 \simeq 3 \times 10^{-2}~. 
\end{align}
Here the electron-sterile mixing, $U_{e4}$, is of interest.  

With the values of mixings and masses given above, and in absence of other exotica, $\nu_s$ should be produced (via $\numu \rightarrow \nu_s$ and  $\nue \rightarrow \nu_s$ oscillations) before BBN with abundance at or close to thermal, so that its contribution to the radiation energy density is comparable to that of the active \ns.  Interestingly, this is compatible with, or even favored by, recent cosmological data.  Roughly, the situation is as follows: \\

\noindent
(i) recent cosmological observations of an excess of radiation, $N_{\rm eff} > 3$, from both the BBN \cite{Cooke:2013cba, Izotov:2013waa} and CMB data \cite{Hou:2012xq, Hinshaw:2012aka, Sievers:2013ica}, which therefore further support the indication of the existence of $\nu_s$. 
\noindent
(ii)  The measurement of the Hubble constant by Planck \cite{Ade:2013zuv} is at tension with the local $H_0$ data \cite{Riess:2011yx}.  
\noindent
(iii)  The measurement of tensor perturbations by BICEP2 \cite{Ade:2014xna}, is at tension with bounds on tensors from Planck's CMB temperature data \cite{Ade:2013zuv}. 

It has been argued very recently that including a sterile neutrino yielding $\sum_j m_{\nu_j} \sim 0.5 \eV$ and $\Delta N_{\rm eff} \sim 0.96 $ can resolve both the tensions at (ii)  \cite{Battye:2013xqa,Hu:2014qma} and (iii) \cite{Zhang:2014dxk,Dvorkin:2014lea,Zhang:2014nta,Li:2014cka,Archidiacono:2014apa, Giusarma:2014zza}. It has to be noted, however, that data lends themselves to multiple interpretations and the situation is still evolving at this time (see \eg, Ref.~\cite{Leistedt:2014sia} for a different view).  

The signature of $\nu_4$ at a tritium neutrino capture experiment is a line displaced by
\begin{align}\label{eq:sterile_peak}
	\Delta K_{e} = m_4 + m_{\nu}
\end{align}
above the endpoint of the beta decay spectrum [see \eref{kej}] (see also Ref.~\cite{Li:2010sn}).  
The detection rate is proportional to the local number density of sterile neutrinos, $n(\nu_s)$, and to the appropriate mixing factor, $|U_{e4}|^2$.  
Let us consider a basic scenario in which $\nu_s$ is produced via oscillations, in absence of other exotica, and accounts for the entire excess of radiation, $\Delta N_{\rm eff} = N_{\rm eff}  - 3.046$.  It can be shown (see, \eg, \cite{Dodelson:1993je,Jacques:2013xr})  that  its momentum distribution is the same as the one of the active neutrinos, up to a constant scaling factor, and therefore the local number density of $\nu_4$ is  \cite{Jacques:2013xr}
\be\label{nnus}
	n(\nu_s) \simeq f_{\rm c} \, n_{0} \, \Delta N_{\rm eff} \com
\ee
where $f_{\rm c} \lesssim 50$ \cite{Ringwald:2004np} is the enhancement factor due to gravitational clustering (see \sref{sec:clustering} and \tref{tabactive}). 

Thus, the ratio of the $\nu_4$  capture rate to the \cnb\ active (Majorana) neutrino capture rate is 
\be
\frac{\Gamma_{\nu_4}}{\Gamma_{\CNB}^{\rm M}} \approx 0.6 ~ \Delta N_{\rm eff} \left( \frac{ |U_{e4}|^2 }{3 \times 10^{-2} } \right) \left( \frac{ f_{\rm c} }{20} \right) \com
\label{gammanus}
\ee
or $\Gamma_{\nu_4} \approx 4.9 \, {\rm yr}^{-1}$. The result in \eref{gammanus} refers to rather optimistic parameters, and therefore should be considered as the best case scenario.  
Although the rate is smaller than for the active  species, its significance in the detector might be boosted by its larger separation from the the endpoint of the beta decay spectrum. The reason is twofold: first, the excess due to $\nu_4$ would be more easily resolved, even with a worse resolution than PTOLEMY; second, the region near the $\nu_4$ peak would be nearly background-free, since the beta decay spectrum falls exponentially with energy.  These aspects  are illustrated in \fref{fig:cartoon}.

\subsection{keV-scale warm dark matter sterile neutrinos}

The above discussion carries over for a sterile neutrino in the keV mass range (see Ref.~\cite{Li:2010vy}), which is a candidate for warm dark matter, and has  number of interesting manifestations depending on its mixing with the active species. The strongest constraints on $U_{e4}$ in this mass range are
\be\label{kevparam}
	 |U_{e4}|^2  \lesssim {\mathcal O}(10^{-9}) ~;
\ee
they come from bounds on the abundance of $\nu_s$ in the early universe, and specifically from data on the spectrum of Large Scale Structures, on observations of the Lyman-$\alpha$ forest, and from X-ray observations constraining $\nu_4$ radiative decay (see \eg, \cite{Smirnov:2006bu,Kusenko:2009up} and references therein). Besides bounds, there are positive claims hinting at the existence of a keV-scale $\nu_s$.  Recently a $3.5 \keV$ X-ray line has been identified in various galaxy clusters \cite{Bulbul:2014sua,Boyarsky:2014jta}.  Interpreting this line with a decaying sterile neutrino state yields the parameters $m_4 \simeq 7~{\rm keV}$ and mixing $\sin^2 2 \theta = 4 |U_{\alpha 4}|^2 \simeq (2-20) \times 10^{-11}$ \cite{Bulbul:2014sua,Boyarsky:2014jta}. Such small mixing values will lead to a corresponding suppression of the neutrino detection rate at PTOLEMY.  However, this suppression is partially offset by an enhancement:  with its larger mass, $\nu_4$ can cluster much more efficiently, and therefore its local abundance could be much larger than the unclustered \cnb\ abundance. Specifically,  if we assume that $\nu_4$ account for 100\% of the dark matter local density, $\rho_{\rm DM} \simeq 0.3~{\rm GeV~ cm^{-3}}$ \cite{Bovy:2012tw}, the clustering enhancement factor is: 
\be
	f_{\rm c} \approx \frac{\rho_{DM} / m_{4}}{n_{0}} \simeq  7.6 \times 10^2 \left( \frac{7 \keV}{m_4} \right) \com
\ee
Taking both the mixing suppression and the clustering enhancement into account, the expected rate at PTOLEMY is given by
\be
	\frac{\Gamma_{\nu_{\rm 4}}}{\Gamma_{\CNB}^{\rm M}} \simeq |U_{e4}|^2  f_{\rm c}  \simeq  7.6 \times 10^{-9} \left( \frac{|U_{e4}|^2}{10^{-11}}\right)\left( \frac{7~{\rm keV}}{m_4}\right)~. 
\label{ratekev}
\ee
Thus, we conclude that the interesting region of the parameter space is out of reach of this type of experiment, although interesting, complementary bounds on $\nu_s$ could be obtained \cite{Betts:2013uya}.


\section{Sensitivity to other non-standard neutrino physics}\label{sec:exotica}

We now turn to other possible effects that might enhance or suppress the \cnb\ capture signal, such as the lepton asymmetry in the neutrino sector, neutrino decay, and the entropy injection after the neutrino decoupling. 

\subsection{Lepton asymmetry}
\label{sub:lepton_asym}

It is established that the universe possesses a cosmic baryon asymmetry, defined as the difference between the number density of baryons and that of anti-baryons: $n_B = n_b - n_{\bar b}$.  Normalized to the photon density, the asymmetry is  $n_B / n_{\gamma} \approx 10^{-10}$ \cite{Ade:2013zuv}.  A neutrino asymmetry, $n_L = n_\nu - n_{\bar \nu}$, is also expected in many models of baryogenesis.  In most models it is expected to be comparable to $n_B$, however there are cases (\eg, \cite{Shi:1998km,Laine:2008pg,Boyarsky:2009ix}), where $O(10^{-3})-O(1)$ lepton asymmetry in the neutrino sector can be created, and the current constraints are at the level of $n_L / n_{\gamma} \lesssim 0.1-0.5$ \cite{Schwarz:2012yw}.  

In \erefs{eq:D_state}{eq:M_state} we enumerated the degrees of freedom for Dirac and Majorana neutrinos.  
An asymmetry may arise between states which are CP conjugates to one another.  If the neutrinos are Dirac particles, then this asymmetry is manifest as $n(\nu_{h_L}) \neq n(\bar{\nu}_{h_R})$, and is conserved in the absence of lepton-number violating interactions.  In the Majorana case, the asymmetry means  $n(\nu_{h_L}) \neq n(\nu_{h_R})$, and  is approximately conserved as long as the helicity-flipping rate is smaller than the Hubble expansion rate \cite{Langacker:1982fg}.  
As discussed in \sref{sec:cnb_evolution}, this is the case for free-streaming neutrinos.  

Let us start by considering the Dirac case, and generalize the neutrino distribution function, \eref{eq:f_fo}, to include an asymmetry. We will assume that each of the three mass eigenstates carries the same asymmetry, because equilibration of flavor is generally expected due to oscillations (see \eg, \cite{Lunardini:2000fy,Dolgov:2002ab}). 
Let $\mu_{\nu}$ be the chemical potential and $\xi_{\nu} = \mu_{\nu} / T_{\nu}$.  Then the number density and energy density  of neutrinos are:  
\begin{align}
	\label{eq:n_hL}
	n(\nu_{h_L}) &
	= N_f \int  \frac{d^3 p}{(2 \pi)^{3}} \frac{1}{e^{(p-\mu_{\nu})/T_{\nu}} + 1} 
	\approx \frac{3 N_f \zeta(3)}{4 \pi^2} T_{\nu}^3 + \frac{N_f}{12} \xi_{\nu} T_{\nu}^3 + O(\mu_{\nu}^2 T_{\nu}) \com \\
	\rho(\nu_{h_L})  &
	= N_f \int  \frac{d^3 p}{(2 \pi)^{3}} \frac{p}{e^{(p-\mu_{\nu})/T_{\nu}} + 1} 
	\approx \frac{7 N_f \pi^2}{240} T_{\nu}^4 + \frac{9 N_f \zeta(3)}{4 \pi^2} \xi_{\nu} T_{\nu}^4 + \frac{N_f}{8} \xi_{\nu}^2 T_{\nu}^4 + O(\mu_{\nu}^3 T_{\nu}) \com  \nonumber
\end{align}
where $N_f = 3$ reflects the sum over flavors, $p \equiv | {\bf p}|$, and we have assumed $\xi_{\nu} \ll 1$ in the expansions on the left side.  The corresponding quantities for anti-neutrinos are given by a change of the sign in $\xi_\nu$: $n(\bar{\nu}_{h_R}) = n(\nu_{h_L})|_{\xi_\nu \rightarrow -\xi_\nu}$, etc.

As we saw in \sref{sec:cnb_detection}, only $n(\nu_{h_L})$ is relevant for \cnb\ detection.  
We immediately see that, compared to the symmetric case ($\xi_\nu=0$) $n(\nu_{h_L})$ is enhanced (suppressed) if $\xi_{\nu} > 0$ ($\xi_{\nu} < 0$). 
Therefore, the \cnb\ capture rate will have a corresponding enhancement (suppression) factor:
\be \label{eq:fLA}
	f_{\xi}^{\rm D}
	= \frac{n(\nu_{h_L})}{n(\nu_{h_L})|_{\xi_{\nu}=0}} \simeq  1 + \frac{\pi^2}{9 \zeta(3)} \xi_{\nu} \approx 1 + 0.91 \xi_{\nu}  \per
\ee
For Majorana neutrinos, the calculation proceeds from \eref{eq:n_hL} in a similar way, however here the quantity relevant to \cnb\ detection is the sum $n(\nu_{h_L}) + n(\nu_{h_R})$ [see \eref{eq:Gamma_CNB}].  
Upon summing, the term linear in $\xi_{\nu}$ cancels out, and the enhancement factor in this case is instead 
\be \label{eq:fLA_M}
	f_{\xi}^{\rm M}
	= \frac{n(\nu_{h_L}) + n(\nu_{h_R})}{[ n(\nu_{h_L}) + n(\nu_{h_R}) ]_{\xi_{\nu}=0}} \simeq 1 + \frac{2 \ln 2}{3 \zeta(3)} \xi_{\nu}^2 \approx 1 + 0.38 \xi_{\nu}^2  \com
\ee
therefore, for Majorana neutrinos capture is always enhanced by asymmetry.  

The lepton asymmetry also translates into an additional energy density, 
\begin{align}
	\rho_{\nu}^{\rm tot} = \rho(\nu_{h_L}) + \rho(\bar{\nu}_{h_R}) \approx \frac{7 N_f \pi^2}{120} T_{\nu}^4 + \frac{N_f}{4} \xi_{\nu}^2 T_{\nu}^4 \com
\end{align}
that increases regardless of the sign of $\xi_{\nu}$.  In cosmology the proxy for $\rho_{\nu}^{\rm tot}$ is the commonly quoted effective number of neutrinos, $N_{\rm eff}$ [\eref{cosmobound}]: 
\be
	N_{\rm eff} = N_f \frac{\rho_{\nu}^{\rm tot}}{\rho_{\nu}^{\rm tot}|_{\xi_{\nu}=0}} \approx N_f + \frac{30 N_f}{7 \pi^2} \xi_{\nu}^2 \com
\ee
where we can immediately read the excess due to the asymmetry:
\begin{align}\label{eq:Delta_Neff}
	\Delta N_{\rm eff} = N_{\rm eff} - N_f \simeq \frac{30 N_f}{7 \pi^2} \xi_{\nu}^2 \simeq  1.3~ \xi_{\nu}^2 \per
\end{align}
The bound on $N_{\rm eff}$, \eref{cosmobound}, thus imply a bound on $\xi_\nu$.  Additionally, a strong bound on $\xi_\nu$ arises from constrains on the neutron to proton ratio at BBN.  
\begin{table}[t]
\begin{center}
\begin{tabular}{|c|c|c|c|}
\hline
$\xi_{\nu}$ & $f_{\xi}^{\rm D}$ & $f_{\xi}^{\rm M}$ & $\Delta N_{\rm eff}$ \\
\hline
0.30 & 1.31 & 1.03 & 0.12 \\
0.45 & 1.50 & 1.08 & 0.27\\
0.60 & 1.71 & 1.14 & 0.48\\
0.90 & 2.21 & 1.32 & 1.10\\
\hline
-0.30 & 0.76 & 1.03 & 0.12\\
-0.45 & 0.66 & 1.08 & 0.27\\
-0.60 & 0.57 & 1.14 & 0.48 \\
-0.90 & 0.43 & 1.32 & 1.10\\
\hline
\end{tabular}
\end{center}
\caption{\label{tabfla} 
The Dirac and Majorana capture enhancement factors, $f_{\xi}^{\rm D}$ and $f_{\xi}^{\rm M}$, as well as $\Delta N_{\rm eff}$, for given values of the lepton asymmetry parameter $\xi_{\nu}$.  Results are obtained by exact calculation [phase space integrations on the left hand side in \eref{eq:n_hL}].  
}
\end{table}

Table \ref{tabfla} shows $f_{\xi}^{\rm D} $, $f_{\xi}^{\rm M}$ and $\Delta N_{\rm eff}$ for a set of values of $\xi_\nu$. From this table, and from \eref{eq:Delta_Neff}, we can infer the maximum capture enhancement   allowed by cosmology. The Planck satellite constraint on $\Delta N_{\rm eff}$, \eref{cosmobound}, translates into $|\xi_{\nu}| \lesssim 0.5$. A more careful analysis of CMB data (WMAP9, SPT, and ACT) finds that an anti-neutrino excess is preferred, roughly $-0.4 \lesssim \xi_{\nu} \lesssim 0.2$ \cite{Schwarz:2012yw}, where the exact range depends on the combination of data sets used.    This interval corresponds to $0.6 \lesssim   f_{\xi}^{\rm D}   \lesssim1.2 $ and $ 1.0 \lesssim f_{\xi}^{\rm M}   \lesssim 1.1 $. When the $\tensor[^4]{\rm He}{}$ abundance is folded in, the bound tightens to $-0.091 \lesssim \xi_{\nu} \lesssim 0.051$ \cite{Schwarz:2012yw}, corresponding to a negligible effect on the \n\ capture rate.

\subsection{Neutrino decay}

Being massive and lepton flavor-violating, \ns\ could be unstable. Given the \n\ mass eigenstate $\nu_i$, with proper lifetime $\tau_i$, observational constraints on its decay are usually expressed in terms of lifetime-to-mass ratio, $\tau_i / m_i$  (see, \eg, Ref.~\cite{Beringer:2012zz} for a collection of the current limits).  The best {\it model-independent} constraint derives from the measured supernova neutrino flux in SN1987A \cite{Hirata:1987hu}: 
\begin{align}\label{eq:SN1987a}
	\frac{\tau}{m} > 10^5~{\rm s \cdot eV^{-1}} 
\end{align}
for the mass eigenstates $\nu_{1}$ and $\nu_{2}$.  In order to discuss model-dependent constraints, it is convenient to classify the decay channels as: 
\begin{itemize}
	\item  Radiative, ``visible", decay.  One of the decay products is a photon.  
	\item ``Weak" decay.  One of the decay products is a (lighter) neutrino, and the other products are invisible. For example, it could be that all the neutrinos ultimately decay down to the lightest neutrino species. 
	\item Invisible decay.  The decay products are exotic, non-interacting particles such as sterile neutrinos.  
\end{itemize}
Very strong limits are placed on the radiative decay channel from solar $\nu$ and $\gamma$ fluxes  \cite{Raffelt:1985rj}
\begin{align}\label{eq:radiative_decay}
	\frac{\tau}{m} \gtrsim 7 \times 10^{9} \, {\rm s \cdot eV^{-1} } 
\end{align}
for the $\nu_1 \approx \nu_e$ mass eigenstate. Because visible decay channels are already strongly constrained, we will focus on the weak and invisible decay channels. \\ 

\noindent
{\it (i)  invisible decay.} \\
If a neutrino completely decays into invisible particles, then the expected \cnb\ capture rate will be suppressed or vanish completely depending on the lifetime. For neutrinos with proper lifetime $\tau^0_\nu$, the suppression factor due to the decay of  into invisible particles is (see, e.g., \cite{Baerwald:2012kc})
\be\label{fdecay}
f_{\rm d}^{\rm inv} = e^{-\lambda_\nu},
\ee    
where
\be\label{lambda}
\lambda_\nu = \int \frac{dt}{\tau_\nu} = \int_{0}^{z_{\rm fo}} \frac{dz}{(1+z) H(z) \gamma(z) \tau_\nu^0} \per
\ee
Here $\tau_\nu (z) = \tau_\nu^0 \gamma(z)$ is the Lorentz-dilated lifetime at epoch $z$, $z_{\rm fo} \simeq 6\times 10^{9}$ is the neutrino decoupling epoch, and the Hubble parameter and the Lorentz factor of a neutrino are respectively given by 
\be
H(z) = H_0 \sqrt{\Omega_{\rm r} (1+z)^4 + \Omega_{\rm m} (1+z)^{3} + \Omega_\Lambda} \com \qquad \gamma(z) = \frac{E_\nu}{m_\nu} = \sqrt{\frac{\overline{p}_0^2}{m_\nu^2} (1+z)^2 + 1} \com
\ee
where $H_0 = 67.04~ \rm{km~ s^{-1} ~ Mpc^{-1}}$, $\Omega_{\rm r} = 9.35 \times 10^{-5}$, $\Omega_{\rm m} = 0.3183 $, $\Omega_{\Lambda} = 0.6817 $ \cite{Ade:2013zuv} and $\overline{p}_0$ is the neutrino momentum in the present epoch [\eref{eq:pbar0}]. 

The calculation of $\lambda_\nu$ is greatly simplified by considering that the integral in \eref{lambda} is dominated by the recent epoch, $z \ll 1$, and that for masses of interest here, the \ns\ were already non-relativistic at that time:  $m_\nu \sim 0.1~{\rm eV}\gg  \overline{p}_0$ [\eref{eq:pbar0}]. Thus, one expects  (and the full calculation confirms this) the non-relativistic result $\lambda_\nu \sim t_0 / \tau_\nu^0$, and hence, 
\be
 f_{\rm d}^{\rm inv} \sim e^{-t_0/\tau_\nu^0} \com
 \label{eq:fd}
 \ee
where the age of the universe is $t_0 = 4.36 \times 10^{17}~{\rm s} $.
From \eref{eq:fd} it follows that a detection of the \cnb\ at PTOLEMY, at a rate consistent with the standard value, would place constraints on the invisible decay rate at the order $\tau_{\nu}^{0} \sim t_0$.  
Instead, a significant suppression, resulting in a negative search, could be evidence for \n\ decay implying a upper bound on the lifetime, $\tau_{\nu}^{0} \lesssim t_0$. 

Interestingly, the sensitivity to the lifetime is {\it  not of the usual form $\tau/m_\nu$}: we can really constrain the lifetime regardless of the mass, provided that the mass is in the range of sensitivity of the experiment.  
This is because this decay test is done with non-relativistic neutrinos, a unique aspect of this setup.   
For comparison with currently available limits, however, we can express the sensitivity as 
\be
	\tau/m_\nu \sim t_0/m_\nu \approx 4.36 \times 10^{18} ~{\rm s \cdot eV^{-1}} \left(\frac{0.1~{\rm eV}}{m_\nu} \right) \com
\label{sens}
\ee
which is enormously better than the current model-independent limit,  \eref{eq:SN1987a}, and  competitive with the cosmological limit for radiative decay, \eref{eq:radiative_decay}\footnote{Strong indirect limits are available, see for instance \cite{Beringer:2012zz, Pakvasa:1999ta}.}. In this way, a \cnb\ direct detection experiment would serve as a complementary probe to other astrophysical searches for neutrino decay.  \\

\noindent
{\it (ii) weak decay.} \\
Let us consider the case of complete decay of all the \cnb\ neutrinos down into the lightest mass eigenstate, which is   
$\nu_1$ for NH or $\nu_3$ for IH, see \sref{sub:hierarchy}.   As a consequence, the \n\ population today  is  entirely made of this state, which is therefore three times more abundant than for stable \ns. This means that \eref{eq:Gamma_CNB} should be modified by replacing $\sum_j |U_{ej}|^2 =1 $ with $ 3 |U_{ei}|^2$, where $i=1$ for NH and $3$ for IH.  
The result is that the capture rate is enhanced or suppressed by a factor
\begin{align}
\label{eq:fweak}
	f_{\rm d}^{\rm w} = \frac{3 |U_{ei}|^2}{\sum_{j} |U_{ej}|^2} = \begin{cases}
	3 |U_{e1}|^2 \approx 2.03 
	& {\rm NH} \\ 
	3 |U_{e3}|^2 \approx 0.068 
	& {\rm IH}  \, 
	\end{cases} \per
\end{align}
For the IH case, neutrino weak decay would lead to a null result. On the other hand, detection would be enhanced in the NH case, provided that the detector resolution is good enough to resolve $m_1$.    The observation of an anomalous rate compatible with \eref{eq:fweak} would result in a lower or upper bound on the \n\ proper lifetime, along the same argument as in the case of invisible decay.  In case of an incomplete decay, the value $f_{\rm d}^{\rm w}$ is intermediate between $1$ and the results in \eref{eq:fweak}.

\subsection{Non-standard thermal history}
\label{sub:nonstandtherm}

The predicted \cnb\ detection rate depends sensitively on the temperature of the relic neutrinos, via the relationship between the temperature and the number density [\eref{eq:n_fo}].  For example, in the Majorana neutrino case our calculated rate is
\be
\Gamma^{\rm M}_{\CNB}  \simeq 8 \, {\rm yr}^{-1} \left( \frac{T_\nu}{1.9^{\circ} \, {\rm K}}\right)^3 \per
\label{tvariable}
\ee
Supposing that new physics were to affect the \cnb\ temperature (while maintaining the thermal distribution), it is immediately clear from \eref{tvariable} that the \cnb\ detection rate could be altered dramatically with even a small temperature change:  for $T_{\nu} \simeq 4^{\circ} \, {\rm K}$ we would have $\Gamma_{\CNB} \simeq 64\, {\rm yr}^{-1}$. Conversely, a colder \cnb\ leads to a smaller capture rate.  

Needless to say, the \cnb\ temperature has never been directly measured.  Its value is predicted to be $T_{\nu} =T^{\rm std}_{\nu} \simeq 1.9^{\circ} \, {\rm K}$ using the observed temperature of the CMB, $T_{\gamma} \simeq 2.7^{\circ} \, {\rm K}$, and the relationship between $T_\nu$ and $T_\gamma$ [see \eref{eq:Tg_of_t}]:
\begin{align}\label{eq:Tnu_ov_Tg}
\frac{T_{\nu}}{T_{\gamma}} = \frac{g_{\ast}^{1/3}(0)}{g_{\ast }^{1/3}(z_{\rm fo})} = \left(\frac{4}{11}  \right)^{1/3} \com
\end{align}
Here $g_{\ast}(z)$ is the effective number of relativistic species. After neutrino freeze out, the plasma consisted of electrons, positrons, and photons giving $g_{\ast}(z_{\rm fo}) = 2 + (7/8) 4 = 11/2$.  After $e^+ e^-$ annihilation all the the entropy is transferred to the photons for which $g_{\ast}(0) = 2$.  

It is possible that  the \cnb\ temperature could be substantially different than $T_{\nu}$ if the thermal history of the universe were modified.  Specifically, we will suppose that physics beyond the Standard Model is responsible for 
an entropy injection. For example, in analogy with the $e^+ e^-$ annihilation scenario, we can consider a new species of particle that is initially coupled to the plasma but decouples and transfers its entropy to the remaining thermalized species.  Alternatively, the entropy injection could arise from an out-of-equilibrium decay or a first order phase transition. If the injection occurs before neutrino decoupling, then both the photons and the neutrinos are heated.  This delays neutrino decoupling, but once the neutrinos have frozen out, the ratio $T_{\nu} / T_{\gamma}$ is unaffected; it is still controlled by $e^+ e^-$ annihilation.  

Next suppose that entropy is injected into the photons after neutrino decoupling but before recombination.  This heats the photons, which must cool for a longer time to reach the measured value of $2.7^{\circ} \, {\rm K}$, and causes the neutrinos to be relatively colder.  The \cnb\ temperature is calculated using \eref{eq:Tnu_ov_Tg} where $g_{\ast}(0) = 2$ and $g_{\ast}(z_{\rm fo}) = 11/2 + \Delta g$ where $\Delta g$ counts the additional degrees of freedom that were in equilibrium prior to the entropy injection.  For instance, if the entropy arises from the freeze out of a single Dirac species then $\Delta g = (7/8)4$ and $T_{\nu} / T_{\gamma} = (2/9)^{1/3}$.  This implies a colder \cnb, $T_{\nu} \simeq 1.6^{\circ} \, {\rm K}$, and a lower \cnb\ capture rate, $\Gamma_{\CNB} \simeq 5 \yr^{-1}$.  

It seems unlikely that an entropy injection could result in a heating of the \cnb\ neutrinos.  Even if the species that freezes out decays into neutrinos (see, \eg, \cite{Kanzaki:2007pd}), this will not increase the \cnb\ temperature, but instead it will lead to a non-thermal spectrum, since the neutrinos are already free streaming.  

A constraint on the \cnb\ temperature, and therefore on entropy injection, arises from  the measurement of $N_{\rm eff} \simeq 3$ from the CMB.  Recall that $N_{\rm eff}$ gives the energy density of relativistic species at the surface of last scattering normalized to the expected \cnb\ temperature.  In the standard thermal history, the \cnb\ temperature is equal to $T_{\nu}^{\rm std}$ at the surface of last scattering, and the neutrinos contribute $N_{\rm eff} \simeq 3$.  
If the neutrinos had a non-standard temperature $T_{\nu} < T_{\nu}^{\rm std}$ then their contribution is suppressed as $N_{\rm eff} \simeq 3 (T_{\nu} / T_{\nu}^{\rm std})^4$.  The Planck measurement of $N_{\rm eff}$, \eref{cosmobound}, translates into the interval $ 1.95^{\circ} \, {\rm K} <   T_{\nu} < 2.03^{\circ} \, {\rm K}$. To allow a larger deviation  of $T_\nu$ from the standard value, one would have to introduce new relativistic degrees of freedom with just the right energy density to compensate for the energy lost by considering the colder \cnb.


\section{Discussion}
\label{sec:disc}

The detection of the \cnb\ via capture on tritium is conceptually interesting, and, for the first time, possibly realistic.  
The existence of a specific experimental proposal, PTOLEMY, motivates the present study on the phenomenology of this technique.  The planned active mass of PTOLEMY is 100 g of tritium, for which the predicted rate is $\Gamma \simeq (4-8) \, \yr^{-1}$.  

Some of the major challenges for a \cnb\ capture experiment are the energy resolution and the background control.  
The signal (if any) due to the \cnb\  will partially overlap with the background from beta decay, and it is reasonable to expect that the signal and the  background might be comparable.  The estimated energy resolution at PTOLEMY will be $\Delta \sim 0.15 \eV$; if the neutrino masses are on the order of $0.07 \eV$ close to the upper limit allowed by cosmology [\eref{cosmobound}], then this resolution is nearly enough to distinguish the signal from the background [\eref{eq:Delta_bound}], but it is not sufficient if the \ns\ are substantially lighter, in the hierarchical spectrum regime.  
Since PTOLEMY will probe only a portion of the parameter space, it is not guaranteed to succeed.   Still, it will represent an important first step towards the development of more sophisticated technologies for \cnb\ capture. 

The spirit of our study is to address the question of what fundamental physics can be learned from a \cnb\ capture experiment, with emphasis on PTOLEMY, but an open mind towards even more ambitious possibilities.   
Below, the main results of our study are summarized. 

\begin{enumerate}

\item  
For 100 grams of tritium, the \cnb\ capture rate is found to be $\Gamma_{\CNB}^{\rm D} \simeq 4 \, \yr^{-1}$ for Dirac neutrinos and $\Gamma_{\CNB}^{\rm M} \simeq 8 \, \yr^{-1}$ for Majorana neutrinos [\eref{eq:signal_rates}].  
This confirms previous calculations \cite{Cocco:2007za, Lazauskas:2007da} where the rate was also found to be $8 \, \yr^{-1}$, although without distinguishing the nature of the neutrinos or working with the polarized capture cross section [see below \eref{eq:sigma_relativistic}], as we have done here. This relative factor of $2$ between the Dirac and Majorana cases has to be taken into account when planning an experimental setup, as it could spell the difference between an indication of the \cnb\ and its discovery. 

\item  
A \cnb\ capture experiment will probe {\it non-relativistic} neutrinos.  
This kinematical regime is completely unexplored at this time, and may reveal interesting properties that are not accessible in the ordinary relativistic regime, such as the distinction between Dirac and Majorana nature of neutrinos, as we discussed above.  This is in striking contrast with the smallness of corrections at the relativistic regime \cite{Zralek:1997sa, Czakon:1999cd}. In principle, the PTOLEMY concept combines two very attractive features that are traditionally separated: the kinematic measurement of the \n\ mass from nuclear decays
(which is relatively well-understood but insensitive to the origin of \n\ mass) 
and the ability to distinguish between Dirac and Majorana \ns.  
The latter so far has been an exclusive feature of neutrinoless double beta decay \cite{Arnaboldi:2008ds, Alessandria:2011rc,Agostini:2013mzu}.  

\item 
The $m_4 \sim 1 \eV$ sterile \n\ favored by MiniBooNE and other oscillation searches could appear at PTOLEMY with a remarkably clean and unambiguous signature: 
an excess of up to $5$ events per year [\eref{gammanus}] 
with an electron kinetic energy that should easily be distinguished from that caused by the active neutrinos and the beta decay endpoint [\eref{eq:sterile_peak}].  
In absence of other exotica, this neutrino should be produced copiously (at or close to thermal abundance) in the early universe.  
This detection could completely resolve the confused situation that we have inherited from oscillation experiments, where different searches lead to conflicting results and open questions exist on systematic uncertainties and parameter degeneracies.  
Additionally, it could also help to resolve the tension between cosmological probes of neutrinos, specifically measurements of $N_{\rm eff}$, as well as resolving the tension between B-mode polarization data from BICEP2 and the Planck bound.  It should be noted, though, that the absence of an excess at the eV-scale would not exclude the LSND / MiniBooNE sterile \n, but instead restrict the allowed region of $|U_{e4}|$.  
The $\keV$-scale sterile neutrinos require much smaller mixing angles if they are to be the dark matter, and this implies a correspondingly small capture rate [\eref{ratekev}].  

\item 
A direct detection of the \cnb\ would be a unique probe of what happened to \ns\ since the CMB decoupling time.  
In principle, it can vastly improve constraints on \n\ decay, and a detection of the \cnb\ would imply a \n\ lifetime longer than the age of the universe [\eref{sens}].  Interestingly, this bound would be on the \n\ lifetime itself, and not on the ratio of lifetime and mass that is probed with relativistic \ns.  A direct detection would also provide the unique opportunity to probe the coupling of \ns\ to gravity through the local \n\ overdensity, and thereby explore late-time phenomena such as \n\ clustering.  Since the \cnb\ capture rate goes like the third power of the \cnb\ temperature, direct detection may be used to test non-standard thermal histories in which the neutrinos are heated or cooled by a late-time entropy injection.  

\item 
We have found that many of the variants on standard neutrino physics lead to enhancements or suppressions of the \cnb\ capture rate.  These include gravitational clustering [\eref{fclus} and \tref{tabactive}], weak decay of neutrinos [\eref{eq:fweak}], the presence of a lepton asymmetry [\erefs{eq:fLA}{eq:fLA_M} and \tref{tabfla}], and a non-standard thermal history [Sec.~\ref{sub:nonstandtherm}].  Certainly, one has to be mindful of uncertainties and degeneracies.  Since a direct detection of the \cnb\ will only provide two pieces of data, the $\nu$ mass scale and the detection rate, it would be impossible to distinguish between different causes of enhancement or suppression of the \cnb, unless the \n\ capture data are combined with the indirect information from cosmological measurements.  

\end{enumerate}

By the time that the PTOLEMY experiment becomes operational, some of the neutrino parameters will hopefully have been measured by other experiments, \eg, the mass hierarchy by accelerator experiments, the mass scale via cosmology and beta decay, and the Dirac or Majorana character via neutrinoless double beta decay.  This information will be a great advantage to PTOLEMY by helping to break the degeneracies in neutrino parameters (discussed above) and thereby allow PTOLEMY to draw more solid conclusions about the physics of the \cnb.  We want to emphasize that the detection of the \cnb\ will not only be a boon to the field of neutrino physics and cosmology, but also could lead to interesting and unexpected new physics that could manifest itself in the regime where \ns\ are non-relativistic.  Therefore, the agenda for PTOLEMY and similar proposals might become richer than previously considered.


\acknowledgments

We are grateful to D. Chung, R. Gran, G. Mangano, M. Messina, C. Quigg and C. Tully for very useful discussions.  
C.L. and E.S. acknowledge the National Science Foundation grant number PHY-1205745. 
A.J.L. is supported by the DOE under Grant No.~DE-SC0008016.


\appendix
\section{Amplitude and cross section for polarized neutrinos}
\label{app:nucapture}

\subsection{Kinematics}
\label{app:Kinematics}

In this section, we present the kinematic relations that arise in the calculations of tritium beta decay and neutrino capture on tritium.  The calculation treats only the nuclear process (masses $m_{\tritium}$ and $m_{\Hethree}$ are nuclear masses), and we comment on the role of the atomic electron at the end.  We work in the rest frame of the tritium nucleus.  

Since tritium beta decay is a three-body process, $\tritium \to \Hethree + e^{-} + \overline{\nu}_j$, the electron can be emitted with a range of momenta.  The maximum possible momentum is obtained when the electron is emitted anti-parallel to both the helium-3 nucleus and the neutrino.  This momentum demarcates the beta decay endpoint, 
\begin{align}\label{app:pend}
	p_{\rm end} = \frac{1}{2 m_{\tritium}} \sqrt{m_{\tritium}^2 - (m_{\Hethree} + m_{\nu} + m_{e})^2} \sqrt{m_{\tritium}^2 - (m_{\Hethree} + m_{\nu} - m_{e})^2} \per
\end{align}
The corresponding electron kinetic energy is given by $K_{\rm end} = \sqrt{p_{\rm end}^2 + m_{e}^2} - m_{e}$ or
\begin{align}\label{app:Kend}
	K_{\rm end} = \frac{(m_{\tritium}-m_e)^2 - (m_{\Hethree}+m_{\nu})^2}{2m_{\tritium}} \per
\end{align}
It is convenient to introduce the Q-value, defined by\footnote{Alternatively, the Q-value may be defined as $\bar{Q} = M_{\tritium} - M_{\Hethree}$ where these are the {\it atomic} masses of tritium and helium-3.  The difference between $Q$ and $\bar{Q}$ is the $O(10 \eV)$ atomic binding energies.  }
\begin{align}\label{app:Q_def}
	Q \equiv m_{\tritium} - m_{\Hethree} - m_{e} - m_{\nu} \com
\end{align}
which corresponds to the total kinetic energy carried away by all three decay products.  Note that no single particle can have a kinetic energy equal to $Q$ because this would require a violation of momentum conservation, that it, it would neglect the recoil of the other decay products.  In terms of $Q$, the endpoint energy can be written 
\begin{align}\label{app:Kend_of_Q}
	K_{\rm end} = Q - K_{\rm recoil}
\end{align}
where
\begin{align}
	K_{\rm recoil} = \frac{m_e}{m_{\tritium}} Q + \frac{Q^2}{2 m_{\tritium}}
\end{align}
is the amount of kinetic energy unavailable to the electron, because it goes into the recoil of the helium-3 nucleus and the neutrino.  It is also convenient to identify the energy 
\begin{align}
	K_{\rm end}^{0} = \frac{(m_{\tritium}-m_e)^2 - m_{\Hethree}^2}{2m_{\tritium}} \com
\end{align}
which is where the endpoint would be located if the neutrino were massless.  

Next let us consider the kinematics of the neutrino capture process, $\nu_j + \tritium \to \Hethree + e^{-}$.  The calculation simplifies greatly if we neglect the momentum of the incident neutrino.  For typical \cnb\ neutrinos, which have a momentum $\overline{p}_0 \approx 6 \times 10^{-4} \eV$ and a mass $m_{\nu} \approx 0.1 \eV$, this is a very good approximation.  A simple calculation gives the kinetic energy of the emitted electron to be 
\begin{align}\label{app:Kcnb}
	K_e^{\text{\cnb}} = \frac{( m_{\tritium} - m_{e} + m_{\nu} )^2 - m_{\Hethree}^2}{2 (m_{\tritium} + m_{\nu}) } \per
\end{align}
This is the energy at which the \cnb\ signal will be located.  Its displacement above the beta decay endpoint is given by
\begin{align}\label{app:DeltaK_def}
	\Delta K = K_e^{\text{\cnb}} - K_{\rm end} = \frac{(m_{\tritium} + m_{\Hethree} + m_{\nu})^2 - m_e^2 }{2 m_{\tritium} (m_{\tritium} + m_{\nu})} \, m_{\nu} \per
\end{align}
If we now make the well-justified approximations $m_{\tritium} \approx m_{\Hethree} \gg m_{e} \gg m_{\nu}$, we arrive at the simple result $\Delta K \approx 2 m_{\nu}$ or equivalently,
\begin{align}\label{app:DeltaK}
	K_e^{\text{\cnb}} \approx K_{\rm end} + 2 m_{\nu} \per
\end{align}
Since $m_{\nu} \gg \overline{p}_0$, we were justified in dropping the neutrino momentum at the start.  

We have focused here on the kinematics of the nuclear processes, but the system we are really interested in is a neutral tritium atom converting into a helium ion.  The energy of the emitted electron, however, should be insensitive to the presence of an atomic cloud, both in the beta decay and neutrino capture process.  
The nuclear process occurs on a short time scale, and on a much longer time scale the bound electron finds itself in an excited state of the helium atom.  The helium ion relaxes to its ground state by emitting a photon.  
For this reason, one should not calculate the kinematics using the atomic states; the photon energy must be included as well, and this approach makes the calculation unnecessarily complicated.   

We will conclude this appendix by numerically evaluating the kinematical variables using the measured masses.  
Although it is not necessary to perform this exercise, since \eref{app:DeltaK} depends only on the neutrino mass, we feel that it is illustrative to the reader.  The nuclear masses of tritium and helium-3 are not provided directly in the {\rm AME}2003 tables \cite{Audi:2002rp}.  Instead they must be derived from the atomic masses, which are 
\begin{align}
	M_{\tritium}({\rm atomic}) & = 3 \, 016 \, 049 \, . \, 2777(25) \, \mu {\rm u} \\
	M_{\Hethree}({\rm atomic}) & = 3 \, 016 \, 029 \, . \, 3191(26) \, \mu {\rm u}
\end{align}
where ${\rm u} = 931.4940090(71) \MeV$.  
The nuclear masses are then calculated as 
\begin{align}\label{app:atom_to_nucl}
	m_{\tritium}({\rm nuclear}) & = M_{\tritium}({\rm atomic}) - m_e + 13.59811 \eV \\
	m_{\Hethree}({\rm nuclear}) & = M_{\Hethree}({\rm atomic}) - 2 m_e + 24.58678 \eV 
\end{align}
where the last term on each line is the atomic binding energy \cite{RevModPhys.39.125}.  The parenthetical values show the $1 \sigma$ errors, and the binding energies have negligible error.  Taking also the measured electron mass from \rref{Beringer:2012zz} we have 
\begin{align}
	m_{e} & \approx 510 \, . \, 998 \, 910(13) \keV \\
	m_{\tritium} & \approx 2 \, 808 \, 920 . \, 8205 (23) \keV \\
	m_{\Hethree} & \approx 2 \, 808 \, 391 . \, 2193 (24) \keV
\end{align}
and
\begin{align}
	Q & \approx 18.6023(34) \keV - m_{\nu} \\
	K_{\rm recoil} & \approx 0.003445729(86) \keV + O(m_{\nu}^2) \\
	K_{\rm end}^0 & \approx 18.5988(34) \keV \\
	K_{\rm end} & \approx 18.5988(34) \keV - m_{\nu} \\
	K_e^{\text{\cnb}} & \approx 18.5988(34) \keV + m_{\nu} \label{kecnb}\\
	\Delta K & \approx 2 m_{\nu} \per
\end{align}
The error is dominated by the uncertainty in the atomic masses.  
Although the error bars on $K_{\rm end}$ and $K_e^{\text{\cnb}}$ are on the order of $3.4 \eV$, and therefore much larger than the neutrino mass, the displacement $\Delta K$ is insensitive to these uncertainties.  

\subsection{The polarized neutrino capture amplitude}
\label{eq:Appendix1}

Here we provide some of the details behind the cross section calculation in \sref{sec:cnb_detection}.  To our knowledge the literature does not contain an explicit calculation of the polarized neutrino capture cross section for this process.  

Starting with the matrix element in \eref{eq:matrix_element}, we first calculate the modulus
\begin{align}
	\abs{ \mathcal{M} }^{2}
	& = \frac{G_F^2}{2} |V_{ud}|^2 |U^*_{ej}|^2 \mathcal{T}_{1}^{\alpha \gamma} \mathcal{T}_{2}^{\beta \delta} \eta_{\alpha \beta} \eta_{\gamma \delta} \com
\end{align}
where
\begin{align}
	\mathcal{T}_{1}^{\alpha \gamma} & \equiv {\rm Tr} \left[ \gamma^{\alpha} (1 - \gamma^5) u_{\nu} 
	\overline{u}_{\nu} \gamma^{\gamma} (1 - \gamma^5) u_{e} \overline{u}_{e}  \right] \\
	\mathcal{T}_{2}^{\beta \delta} & \equiv {\rm Tr} \left[ \gamma^{\beta} \left( f - \gamma^{5} g\right) u_{n} 
	\overline{u}_{n} \gamma^{\delta} \left( f - \gamma^{5} g\right) u_{p} \overline{u}_{p}  \right] \per
\end{align}
To reduce notion clutter, we have dropped the index $j$ that indicates the neutrino mass eigenstate.  As described in the text, we will sum the spins of the final state electron and proton, and we will average the spin of the initial state neutron.  Doing so gives
\begin{align}\label{app:Msq}
	\overline{\abs{\mathcal{M}}^2} 
	& = \frac{1}{2} \sum_{s_{n}, s_{e}, s_{p} = \pm 1/2} \abs{\mathcal{M}}^{2}
	= \frac{G_F^2}{4} |V_{ud}|^2 |U^*_{ej}|^2 \overline{\mathcal{T}_1^{\alpha \gamma}} \ \overline{\mathcal{T}_2^{\beta \delta}} \eta_{\alpha \beta} \eta_{\gamma \delta} \com
\end{align}
where 
\begin{align}\label{app:Tbar}
	\overline{\mathcal{T}_{1}^{\alpha \gamma}} & = \sum_{s_e = \pm 1/2} {\rm Tr} \left[ \gamma^{\alpha} (1 - \gamma^5) u_{\nu} 
	\overline{u}_{\nu} \gamma^{\gamma} (1 - \gamma^5) u_{e} \overline{u}_{e}  \right] \com \\
	\overline{\mathcal{T}_{2}^{\beta \delta}} & = \sum_{s_n, s_p = \pm 1/2} {\rm Tr} \left[ \gamma^{\beta}  \left( f - \gamma^{5} g\right) u_{n} 
	\overline{u}_{n} \gamma^{\delta} \left( f - \gamma^{5} g\right) u_{p} \overline{u}_{p}  \right] \per
\end{align}
We now require the completeness relations, 
\begin{align}\label{app:completeness}
	& \sum_{s_{i} = \pm 1/2} u_{i} \overline{u}_i = \bigl( \slashed{p}_{i} + M_{i} \bigr)
	\qquad  {\rm for} \quad 
	i = n,p,e \nonumber \com \\
	& u_{\nu} \overline{u}_{\nu} = \frac{1}{2} \bigl( \slashed{p}_{\nu} + M_{\nu} \bigr) \bigl( 1 + 2 s_{\nu} \gamma^5 \slashed{S}_{\nu} \bigr) \com
\end{align}
where
\begin{align}
	(S_{\nu})^{\alpha} = \left( \frac{\abs{{\bf p}_{\nu}}}{m_{\nu}} \ , \ \frac{E_{\nu}}{m_{\nu}} \hat{\bf p}_{\nu} \right)
\end{align}
is the neutrino spin vector.  
Inserting \eref{app:completeness} into \eref{app:Tbar} yields 
\begin{align}
	\overline{\mathcal{T}_{1}^{\alpha \gamma}} & = \frac{1}{2}  {\rm Tr} \left[ \gamma^{\alpha} (1 - \gamma^5) \bigl( \slashed{p}_{\nu} + m_{\nu_j} \bigr) \bigl( 1 + 2 s_{\nu} \gamma^5 \slashed{S}_{\nu} \bigr) \gamma^{\gamma} (1 - \gamma^5) \bigl( \slashed{p}_e + m_e \bigr) \right] \\
	\overline{\mathcal{T}_{2}^{\beta \delta}} & = {\rm Tr} \left[ \gamma^{\beta} \bigl( f - \gamma^{5} \bigr) \bigl( \slashed{p}_n + m_n \bigr) \gamma^{\delta} \bigl( f - \gamma^{5} g\bigr) \bigl( \slashed{p}_p + m_p \bigr) \right] \per
\end{align}
The traces are evaluated using the Mathematica package ``Tracer'' \cite{Jamin:1991dp}, and we find 
\begin{align}
	\overline{\mathcal{T}_1^{\alpha \gamma}} \ \overline{\mathcal{T}_2^{\beta \delta}} \eta_{\alpha \beta} \eta_{\gamma \delta} = 32 \Biggl\{ &
	(g+f)^2 \Bigl[ (p_e \cdot p_p) (p_{\nu} \cdot p_n) \Bigr] 
	+ (g-f)^2 \Bigl[ (p_e \cdot p_n) (p_{\nu} \cdot p_{p}) \Bigr] \nonumber \\
	& + ( g^2 - f^2 ) \Bigl[ m_n m_p (p_e \cdot p_{\nu}) \Bigr] 
	\Biggr\} 
	- 2 s_{\nu} m_{\nu_j} \Biggl\{
	(g+f)^2 \Bigl[ (p_{e} \cdot p_{p}) (S_{\nu} \cdot p_{n}) \Bigr] \nonumber \\
	& + (g-f)^2 \Bigl[ (p_{e} \cdot p_{n}) (S_{\nu} \cdot p_{p}) \Bigr] 
	+ (g^2 - f^2) \Bigl[ m_n m_p (p_{e} \cdot S_{\nu}) \Bigr] 
	\Biggr\} \per 
\end{align}
The spin-independent terms ($s_{\nu} = 0$) match with \rref{Lazauskas:2007da}, and the spin-dependent terms are new.  

We now specify to the rest frame of the neutron (parent nucleus) where 
\begin{align}\label{eq:lab_frame}
	(p_{n})^{\mu} = \left\{ m_{n} ; {\bf 0} \right\}
	\quad , \quad
	(p_{\nu})^{\mu} = \left\{ E_{\nu} ; {\bf p}_{\nu} \right\}
	\quad , \quad
	(p_{p})^{\mu} = \left\{ E_{p} ; {\bf p}_{p} \right\}
	\quad , \quad
	(p_{e})^{\mu} = \left\{ E_{e} ; {\bf p}_{e} \right\} \per
\end{align}
Neglecting the proton (daughter nucleus) recoil, ${\bf p}_p \ll m_{p}$, we obtain 
\begin{align}\label{app:Tbar_final}
\overline{\mathcal{T}_1^{\alpha \gamma}} \ \overline{\mathcal{T}_2^{\beta \delta}} \eta_{\alpha \beta} \eta_{\gamma \delta} =	\ 
	& 32 m_n m_p E_e E_{\nu} \Biggl\{
	2(g^2 + f^2) 
	+ ( g^2 - f^2 ) \Bigl[ 1 - \frac{{\bf p}_e}{E_e} \cdot \frac{{\bf p}_{\nu}}{E_{\nu}} \Bigr] 
	\Biggr\} \nn \\
	& - 2 s_{\nu}  m_{n} m_{p} E_{e} \abs{{\bf p}_{\nu}} \Biggl\{
	 2(g^2 + f^2)
	+ (g^2 - f^2) \Bigl[ 1 - \frac{E_{\nu}}{\abs{{\bf p}_{\nu}}} \frac{{\bf p}_{e}}{E_e} \cdot \frac{{\bf p}_{\nu}}{\abs{{\bf p}_{\nu}}} \Bigr] 
	\Biggr\} \nn \\
	= \ & 32 m_n m_p E_e E_{\nu} \Bigl[ 
	(f^2 + 3 g^2) (1 - 2 s_{\nu} v_{\nu} )
	+ ( f^2 - g^2) ( v_{\nu} - 2 s_{\nu} ) v_{e} \cos \theta
	\Bigr] \com
\end{align}
where $\cos \theta = {\bf p}_e \cdot {\bf p}_{\nu_j} / ( \abs{{\bf p}_{e}} \abs{{\bf p}_{\nu_j}} )$ and $v_{i} \equiv \abs{{\bf p}_i} / E_i$.  
Inserting \eref{app:Tbar_final} into \eref{app:Msq} gives \eref{eq:squared_amp}.  

In the center of momentum frame the differential cross section is \cite{Beringer:2012zz}
\begin{align}\label{eq:sigma_PDG}
	\frac{d\sigma}{dt} = \frac{1}{64 \pi} \frac{1}{s} \frac{1}{\abs{{\bf p}_{\rm \nu \, cm}}^2} \overline{\abs{\mathcal{M}}^2} \com
\end{align}
where $s = (p_n + p_{\nu})^2$ and $t = (p_e - p_{\nu})^2$.  
The Mandelstam variables can be evaluated in the lab frame using \eref{eq:lab_frame} to find 
\begin{align}
	s & = (m_n + E_{\nu})^2 - \abs{{\bf p}_{\nu}}^2 = m_n^2 + 2 m_n E_{\nu} + m_{\nu}^2 \simeq m_n^2 \com \\
	t & = (E_e - E_{\nu})^2 - \abs{ {\bf p}_e - {\bf p}_{\nu} }^2 \simeq (m_e - m_{\nu})^2 + 2 |{\bf p}_e| |{\bf p}_{\nu}| \cos \theta \com
\end{align}
and $dt / d\cos \theta = 2 |{\bf p}_e| |{\bf p}_{\nu}|$.  
After also making the replacement ${\bf p}_{\rm \nu \, cm} = {\bf p}_{\rm \nu \, lab} (m_n / \sqrt{s}) \simeq {\bf p}_{\nu}$, we obtain
\begin{align}\label{eq:diff_cross_section_appendix}
\frac{d\sigma}{d\cos \theta} = \frac{1}{32 \pi} \frac{1}{m_n^2} \frac{|{\bf p}_e|}{|{\bf p}_{\nu}|} \overline{\abs{\mathcal{M}}^2}  \per
\end{align}


\end{document}